\documentclass[a4paper,11pt]{article}

\usepackage[utf8]{inputenc}
\usepackage{amsmath,amssymb,subcaption}
\usepackage{tikz}
\usepackage[cm]{fullpage}
\usepackage[shortcuts]{extdash}
\usepackage[nosort]{cite}
\usepackage{hyperref}

\allowdisplaybreaks

\usetikzlibrary{arrows.meta}
\tikzset{>={Stealth[length=2mm]}}
\def\Z{{\mathbb Z}}

\def\C{{\mathbb C}}

\def\e{{\rm e}}
\def\sign{\mathop{\rm sign}\nolimits}
\def\const{\mathinner{\rm const}}
\def\Ker{\mathop{\rm Ker}}

\def\sbm#1{{\arraycolsep=.5pt\def\arraystretch{.5}\left[{\tiny\Matrix{#1}}\right]}}

\def\eq#1$$#2$${\begin{equation#1}#2\end{equation#1}}
\def\d{\partial}
\def\bd{\bar\partial}
\def\bb{{\bar b}}
\def\bc{{\bar c}}
\def\bJ{{\bar J}}
\def\bl{{\bar l}}
\def\bz{{\bar z}}
\def\bw{{\bar w}}
\def\bX{{\bar X}}
\def\bY{{\bar Y}}
\def\bgamma{{\bar\gamma}}
\def\bDelta{{\bar\Delta}}
\def\ve{\varepsilon}
\def\bve{{\bar\varepsilon}}
\def\tn{{\tilde n}}

\def\i{{\rm i}}

\def\cF{{\mathcal F}}
\def\bcF{\bar\cF}
\def\bF{{\mathbb F}}
\def\bbF{\bar\bF}

\def\cI{{\mathcal I}}
\def\cN{{\mathcal N}}

\def\cQ{{\mathcal Q}}
\def\cV{{\mathcal V}}
\long\def\subeq#1{\begin{subequations}#1\end{subequations}}
\def\Align#1$$#2$${\begin{align#1}#2\end{align#1}}
\def\Aligned#1{\begin{aligned}{}#1\end{aligned}}
\def\Multline#1$$#2$${\begin{multline#1}#2\end{multline#1}}
\def\Cases#1{\begin{cases}#1\end{cases}}
\def\Gather#1$$#2$${\begin{gather#1}#2\end{gather#1}}
\def\Gathered#1{\begin{gathered}{}#1\end{gathered}}
\def\Matrix#1{\begin{matrix}#1\end{matrix}}
\def\pMatrix#1{\begin{pmatrix}#1\end{pmatrix}}

\def\Cases#1{\begin{cases}#1\end{cases}}

\def\Rp{${\rm R}_+$}
\def\Rm{${\rm R}_-$}
\def\lcolon{\mathopen{\,:}}
\def\rcolon{\mathclose{:\,}}
\def\crosscolon{\substack{\scriptscriptstyle\times\\\scriptscriptstyle\times}}
\def\lcrosscolon{\mathopen{\,\crosscolon}}
\def\rcrosscolon{\mathclose{\crosscolon\,}}
\def\half{{\textstyle{1\over2}}}
\def\Vir{\mathinner{\it Vir}}
\def\vac{\text{vac}}
\def\ctg{\mathop{\rm ctg}\nolimits}

\makeatletter

\def\section{\@startsection{section}{1}{\z@}%
                                   {-3.5ex \@plus -1ex \@minus -.2ex}%
                                   {2.3ex \@plus.2ex}%
                                   {\normalfont\normalsize\bfseries}}
\def\subsection{\@startsection{subsection}{2}{\z@}%
                                     {-3.25ex\@plus -1ex \@minus -.2ex}%
                                     {1.5ex \@plus .2ex}%
                                     {\normalfont\normalsize\bfseries\itshape}}
\def\@seccntformat#1{\csname the#1\endcsname.~~}
\long\def\@makecaption#1#2{%
  \vskip\abovecaptionskip
  \sbox\@tempboxa{\small#1. #2}%
  \ifdim \wd\@tempboxa >0.9\hsize
  {\leftskip=0.05\hsize\rightskip=0.05\hsize\relax\small
    #1. #2\par}
  \else
    \global \@minipagefalse
    \hb@xt@\hsize{\hfil\box\@tempboxa\hfil}%
  \fi
  \vskip\belowcaptionskip}
\def\Appendix{\appendix
  \def\@seccntformat##1{Appendix~\csname the##1\endcsname.~~}}

\let\over\@@over
\let\atop\@@atop
\let\above\@@above
\let\overwithdelims\@@overwithdelims
\let\atopwithdelims\@@atopwithdelims
\let\abovewithdelims\@@abovewithdelims

\makeatother

\numberwithin{equation}{section}

\begin{document}

\title{The free field representation for the $GL(1|1)$ WZW model revisited}

\author{Michael Lashkevich\\[\medskipamount]
\parbox[t]{0.9\textwidth}{\normalsize\itshape\raggedright
Landau Institute for Theoretical Physics, 142432, Chernogolovka, Russia;\medspace%
\footnote{Mailing address.}
\\[\smallskipamount]
Kharkevich Institute for Information Transmission Problems, 19 Bolshoy Karetny per., 127994 Moscow, Russia;
\\[\smallskipamount]
E-mail: lashkevi@landau.ac.ru}}
\date{}

\maketitle

\begin{abstract}
The $GL(1|1)$ WZW model in  the free field realization that uses the $bc$ system is revisited. By bosonizing the $bc$ system we describe the Neveu\--Schwarz and Ramond sector modules $\cV^{\text{NS}}_{en}=\bigoplus_{l\in\Z}\cV^l_{en}$ and $\cV^{\text{R}}_{en}=\bigoplus_{l\in\Z+{1\over2}}\cV^l_{en}$ in terms of the subspaces of a given fermion number~$l$. We show that there are two sectors of mutually local operators, each consists of all Neveu\--Schwarz operators and of Ramond operators with either integer or half\-/integer spins. Conformal blocks and structure constants are found for operators that correspond the highest weight vectors of the spaces $\cV^l_{en}$. The crossing and braiding matrices are considered and the hexagon and pentagon equations are shown to be satisfied for typical modules. The degenerate case of conformal blocks with atypical (logarithmic) modules as intermediate states is considered. The known conformal block decomposition of correlation functions in the degenerate case is shown to be related to the degeneration splitting in the crossing and braiding relations. The scalar product in atypical modules is discussed. The decomposition of unity in the full correlation functions in the degenerate case in terms of this scalar product is explained.
\end{abstract}

%\newpage

\section{Introduction}

The seminal papers \cite{Rozansky:1992rx,Rozansky:1992td} by Rozansky and Saleur started systematic investigation of the $GL(1|1)$ WZW model in particular, and the structure of logarithmic CFTs in general. Bosonization of the $GL(1|1)$ WZW model involving the $bc$ ghost system was proposed, and in more detail this approach was developed later in the paper~\cite{Schomerus:2005bf}. If was realized~\cite{LeClair:2007aj,Creutzig:2008an} that the model can be reformulated as a system of symplectic fermions~\cite{Kausch:2000fu}. This description provides a more direct way to obtain correlation functions and structure constants~\cite{Troost:2017xtj} than the traditional Dotsenko\--Fateev type free field realization~\cite{Dotsenko:1984nm,Dotsenko:1984ad}. Despite this fact we are going to show that the original free field representation is consistent and fully describes the theory on a sphere. Part of our discussion is very close to that performed in~\cite{Santachiara:2013gna} for the logarithmic extensions of the minimal conformal models and the Liouville theory.

The structure of the representation of the algebra $\widehat{sl}(1|1)$ was extensively studied in \cite{Creutzig:2011cu,Creutzig:2011np,Creutzig:2013hma,Babichenko:2020xun,Creutzig:2020zom}. The characters of the representations were calculated and their modular properties were established. In this paper we will consider representations of the algebra $\widehat{gl}(1|1)$ as unions of representations of the Virasoro algebra extended by modes of the Cartan currents of $\widehat{gl}(1|1)$, which is very natural in the free field representation. We will see that the free field representation dictates us that there are, in fact, two versions of the current algebra, of the Neveu\--Schwarz (NS) and of the Ramond (R) type, with two different classes of representation. An important point is that the full quantum field theory contains both NS and R operators. The Ramond operators split into two classes: $\text{R}^+$ of integer spins and $\text{R}^-$ of half\-/integer spins. We study the mutual locality of the operators and find that the NS operators are mutually local with all NS and R operators, while the $\text{R}^+$ operators and the $\text{R}^-$ operators are mutually semilocal. It results in the conclusion that there are two sectors of mutually local operators: $\text{NS}+\text{R}^+$ and $\text{NS}+\text{R}^-$. The crossing relations for the conformal blocks make it possible to establish structure constants for all operators primary with respect to the extended Virasoro algebra in a very similar way as it was done in~\cite{Dotsenko:1984ad} for minimal conformal models. The only difference is that due to continuous spectrum of the $\widehat{sl}(1|1)$ model the structure constants are obtained as solutions to functional rather than algebraic relations.

The continuous spectrum of the $\widehat{sl}(1|1)$ model allows a construction of any atypical (logarithmic) representation as a `limit' of typical ones. In this limit the representation gets a structure of a filtration. But as a vector space it can be split into two components: `normal' and `logarithmic'. Let us symbolically denote `normal' vectors as $|v^\circ\rangle$ and `logarithmic' ones as $|v'\rangle$. We may define the Shapovalov form as a limit of the Shapovalov forms of typical representations. The $\langle v^\circ|v^\circ\rangle$ elements of the Shapovalov form vanish, the $\langle v^\circ|v'\rangle$ and $\langle v'|v^\circ\rangle$ are given by reduced limit of the Shapovalov forms of the typical representations, while those of $\langle v'|v'\rangle$ type are proportional to a parametric derivative of the Shapovalov forms of typical representations. In the full field theory there are two chiral algebras, and the space of states is a sum of tensor products of pairs of such representations. The states (or operators due to the operator\--state correspondence) of the field theory again split into `normal' states $|\Phi^\circ\rangle$ and `logarithmic' states $|\Psi\rangle$. Symbolically we may write the relation between physical states and vectors in the representation as follows: $|\Phi^\circ\rangle\sim|v^\circ\rangle\otimes|v^\circ\rangle$, $|\Psi\rangle\sim|v^\circ\rangle\otimes|v'\rangle+|v'\rangle\otimes|v^\circ\rangle$. The Shapovalov form for the spaces of states are expressed in terms of those for representations and have a similar structure.

Another problem that the `traditional' free field representation helps to answer is that of consistency of the braiding and crossing matrices with the Moore\--Seiberg (MS) relations. Moore and Seiberg~\cite{Moore:1988uz,Moore:1988ss,Moore:1988qv,Moore:1989vd} established that conformal blocks of a reasonable conformal field theory must satisfy a set of consistency relations, which are written in terms of the braiding and crossing matrices. The subtlety of this approach is that the conformal blocks only satisfy the consistency relations in a very special normalization. In the example of the $\widehat{gl}(1|1)$ model we show that there exists such a normalization of conformal blocks, but it is not quite natural in terms of the free field representation. The MS relations are proved for typical representations, but the continuity of the spectrum allows us to continue the crossing and braiding matrices to atypical representations. If we approach a logarithmic point in the intermediate representations, the crossing and braiding matrices become degenerate, and we need to chose another basis of intermediate states to avoid degeneracy in consistency with~\cite{Rozansky:1992td}. The next question to be understood is which physical states are intermediate in the full correlation functions of physical operators at the degenerate points. We construct a `decomposition of unity' in the correlation functions and show that it is expressed in terms of the inverse Shapovalov form. This `decomposition of unity' consists of four types of terms: $|\Phi^\circ\rangle\langle\Phi^\circ|$, $|\Phi^\circ\rangle\langle\Psi|$, $|\Psi\rangle\langle\Phi^\circ|$ and $|\Psi\rangle\langle\Psi|$. We show that only $|\Phi^\circ\rangle\langle\Psi|$ and $|\Psi\rangle\langle\Phi^\circ|$ terms produce logarithmic contributions to correlation functions.

The paper is organized as follows. In Sect.~\ref{sec-free-field} we describe the model and its free field realization. We show how the typical modules are split into submodules according to the fermion number and describe the Neveu\--Schwartz and Ramond sectors. In Sect.~\ref{sec-correlation-functions} we study the four\-/point conformal blocks of the operators corresponding to the highest weight vectors in each submodule, construct the four\-/point correlation functions and calculate the structure constants.  In Sect.~\ref{sec-braiding-crossing} the braiding and crossing matrices are studied and the Moore\--Seiberg pentagon and hexagon equations are checked in the typical case. Special (atypical) modules and the corresponding logarithmic operators are considered in Sect.~\ref{sec-logops}. We also discuss the structure of the Shapovalov form in this limit. In Sect.~\ref{sec-cb-degen} we discuss the atypical modules in the intermediate channel of the conformal blocks. We show that this case corresponds to degeneration of the braiding and crossing matrices. This degeneration is related to the degeneration of the standard basis of conformal blocks at the special points. By choosing an alternative basis, which is nondegenerate at special points, the degeneration of these matrices is eliminated, and the correlation functions possess a well\-/defined conformal block decomposition. We also study the decomposition of unity in the intermediate channel and find the contributions responsible to logarithmic behavior of the correlation functions. In Sect.~\ref{sec-discussion} we briefly summarize the results.

\section{Free field realization of the $\widehat{gl}(1|1)$ current algebra and its typical representation}
\label{sec-free-field}

In this section we recall the structure of current algebra and of its typical representations in the $bc$ free field realization and fix notation.

\subsection{Current algebra and its free field representation}
\label{sec-current-algebra}

In this paper we will consider the affine superalgebra $\widehat{gl}(1|1)$, more precisely, the two versions of this algebra: the Neveu\--Schwarz (NS) version $\widehat{gl}(1|1)_\text{NS}$ and the Ramond (R) version $\widehat{gl}(1|1)_\text{R}$. To write down both of them more uniformly, let $\lambda=0$ for the NS version and $\lambda=1$ for the R version. These algebras are generated by the even elements $J^E_k,J^N_k$ ($k\in\Z$), odd elements $J^+_r,J^-_r$ ($r\in\Z+{\lambda\over2}$) and the central element~$\kappa$.%
\footnote{The definition of NS and R sectors is adjusted here to integer spin odd currents.}
The nonzero commutation relations are given by
\subeq{\label{gl11-CR}
\Align$${}
[J^N_k,J^E_{k'}]
&=k\kappa\delta_{k+{k'},0}
\label{JNJE-CR}
\\
\{J^+_r,J^-_{r'}\}
&=J^E_{r+r'}+r\kappa\delta_{r+r',0}
\label{J+J--CR}
\\
[J^\pm_r,J^N_k]
&=\mp J^\pm_{r+k}.
\label{JpmJN-CR}
$$}
Define the operators $L_k$ ($k\in\Z$) by a kind of the Sugawara construction~\cite{Rozansky:1992rx}
\eq$$
L_k={1\over2\kappa}\sum_{k'\in\Z}\left(2\lcrosscolon J^E_{-k'}J^N_{k+k'}\rcrosscolon+{1\over\kappa}\lcrosscolon J^E_{-k'}J^E_{k+k'}\rcrosscolon\right)
+{1\over2\kappa}\sum_{r\in\Z+{\lambda\over2}}\left(\lcrosscolon J^-_{-r}J^+_{k+r}\rcrosscolon-\lcrosscolon J^+_{-r}J^-_{k+r}\rcrosscolon\right),
\label{T-Sugawara-modes}
$$
where $\lcrosscolon A_rB_{r'}\rcrosscolon=A_rB_{r'}$ for $r'\ge r$ and $\lcrosscolon A_rB_{r'}\rcrosscolon=\pm B_{r'}A_r$ for $r'<r$ (the sign is plus for all cases except that of a product of two odd operators). They form the Virasoro algebra
\eq$$
[L_k,L_{k'}]=(k-k')L_{k+k'}
\label{Lk-commut}
$$
with the zero central charge. The generators can be gathered into currents:
\eq$$
J^{E,N}(z)=\sum_{k\in\Z}J^{E,N}_kz^{-k-1},
\qquad
J^\pm(z)=\sum_{r\in\Z+{\lambda\over2}}J^\pm_rz^{-r-1},
\qquad
T(z)=\sum_{k\in\Z}L_kz^{-k-2},
\label{JnLn-def}
$$
and the commutation relations (\ref{gl11-CR}) can be written as operator product expansion:
\subeq{\label{gl11-OPE}
\Align$$
J^N(z')J^E(z)
&={\kappa\over(z'-z)^2}+O(1),
\label{JNJE-OPE}
\\
J^+(z')J^-(z)
&={\kappa\over(z'-z)^2}+{J^E(z)\over z'-z}+O(1),
\label{J+J--OPE}
\\
J^\pm(z')J^N(z)
&=\mp{J^\pm(z)\over z'-z}+O(1).
\label{JpmJN-OPE}
$$
}
In terms of the currents the Sugawara construction reads
\eq$$
T(z)={1\over2\kappa}
\left(2\lcrosscolon J^E(z)J^N(z)\rcrosscolon+{1\over\kappa}\lcrosscolon\left(J^E(z)\right)^2\rcrosscolon
+\lcrosscolon J^-(z)J^+(z)\rcrosscolon-\lcrosscolon J^+(z)J^-(z)\rcrosscolon\right),
\label{T-Sugawara}
$$
where
$$
\lcrosscolon A(z)B(z)\rcrosscolon=\lim_{\delta\to0}\left(A(z+\delta)B(z-\delta)-\text{(singular part)}\right).
$$
In this form the construction can be used to define a conformal field theory with $T(z)$ being the energy\-/momentum tensor component. Note that in the form of currents the algebras $\widehat{gl}(1|1)_\text{NS}$ and $\widehat{gl}(1|1)_\text{R}$ look the same, but differ by the sets of admissible representations. That is why the NS and R sectors coexist in the same field theory.

The algebra (\ref{gl11-CR}) admits an automorphism $J^E_k\to\Lambda J^E_k$, $J^-_r\to\Lambda J^-_r$, $\kappa\to\Lambda\kappa$ ($\Lambda\ne0$). This automorphism leaves the Virasoro algebra generators $L_k$ unchanged. This means that without loss of generality we may assume
\eq$$
\kappa=1.
\label{kappa1}
$$

Now, following~\cite{Rozansky:1992rx} consider a system of free bosonic fields $X(z)$, $Y(z)$ with the only nonzero pair correlation functions
\eq$$
\langle X(z')Y(z)\rangle=\langle Y(z')X(z)\rangle=\log{1\over z'-z}
\label{XY-pair}
$$
and the energy\-/momentum tensor
\eq$$
T_{XY}(z)=-\lcolon\d X\,\d Y\rcolon-{\i\over2}\,\d^2Y
\label{TXY-def}
$$
and the fermion $bc$ system with the correlation functions
\eq$$
\langle b(z')c(z)\rangle=\langle c(z')b(z)\rangle={1\over z'-z}
\label{bc-pair}
$$
and the energy\-/momentum tensor
\eq$$
T_{bc}=-\lcolon b\,\d c\rcolon.
\label{Tbc-def}
$$
Then the spin~1 currents
\subeq{\label{J-def}
\Align$$
J^N(z)
&=\lcolon bc\rcolon-\i\,\d X+{\i\over2}\,\d Y,
\label{JN-def}
\\
J^E(z)
&=-\i\,\d Y,
\label{JE-def}
\\
J^+(z)
&=b,
\label{J+-def}
\\
J^-(z)
&=-\i c\,\d Y+\d c,
\label{J--def}
$$
}
satisfy the relations~(\ref{gl11-OPE}). The energy\-/momentum tensor (\ref{T-Sugawara}) reduces to the sum of free boson energy\-/momentum tensors:
\eq$$
T(z)=T_{XY}(z)+T_{bc}(z).
\label{T-def}
$$

%%%%%%%%%%%%%%%%%%%%%%%%%%%%%%%%%%%%%%%%%%%%%%%%%%%%%%%%%%%%%%%%%%%%%%%%
\subsection{Vertex operators: the Neveu--Schwarz and Ramond sectors}

Define the vertex operators $\Phi^\pm_{en}(z)$, which depend on two real parameters $e$ and $n$, by means of the OPEs
\subeq{\label{Phipm-OPE}
\Align$$
J^\pm(z')\Phi^\pm_{en}(z)
&=O(1),
\label{Jpm-Phipm-OPE}
\\
J^\mp(z')\Phi^\pm_{en}(z)
&={e^{1\pm1\over2}\Phi^\mp_{en}(z)\over z'-z}+O(1),
\label{Jmp-Phipm-OPE}
\\
J^N(z')\Phi^\pm_{en}(z)
&={(n\pm1/2)\Phi^\pm_{en}(z)\over z'-z}+O(1),
\label{JN-Phipm-OPE}
\\
J^E(z')\Phi^\pm_{en}(z)
&={e\Phi^\pm_{en}(z)\over z'-z}+O(1).
\label{JE-Phipm-OPE}
$$
}
The conformal dimensions of both vertex operators $\Phi^\pm_{en}(z)$ are the same and are given by
\eq$$
\Delta_{en}=e\left(n+{e\over2}\right)=e\left(\tn-{1\over2}\right),
\qquad
\tn=n+{e+1\over2}.
\label{Delta-en-def}
$$
These operators can be represented in terms of the free fields as
\eq$$
\Phi^+_{en}(z)=V_{e\tn}(z)\equiv\lcolon\e^{-\i(eX+\tn Y)(z)}\rcolon,
\qquad
\Phi^-_{en}(z)=c(z)V_{e\tn}(z).
\label{Phipm-def}
$$
The vector $|v^+_{en}\rangle=\Phi^+_{en}(0)|\vac\rangle$ is the highest weight vector:
\eq$$
J^E_r|v^+_{en}\rangle=0,\quad J^N_r|v^+_{en}\rangle=0,\quad J^+_{r-1}|v^+_{en}\rangle=0,\quad J^-_r|v^+_{en}\rangle=0
\quad\forall\ r\in\Z,\ r>0.
\label{NS-hwv-def}
$$
The operators $J^E_r,J^N_r,J^+_{r-1},J^-_r$ with $r\le0$ generate the Verma module $\cV^\text{NS}_{en}$ by acting on this vacuum. For non\-/integer values of $e$ these module are irreducible. We defer the discussion of atypical modules, which appear at integer values of $e$, to Section~\ref{sec-braiding-crossing}.

Now extend the system of vertex operators. To do it use the bosonized form of the $bc$ system:
\eq$$
b(z)=\lcolon\e^{-\gamma-\i\chi(z)}\rcolon,
\qquad
c(z)=\lcolon\e^{\gamma+\i\chi(z)}\rcolon,
\label{bc-chi}
$$
where $\chi(z)$ is a free boson field with the pair correlation function
\eq$$
\langle\chi(z')\chi(z)\rangle=\log{1\over z'-z}
\label{chichi-corr}
$$
and $\gamma$ is an algebraic element. We will assume that
\eq$$
\langle\e^{a\gamma}\rangle=0,
\quad\text{if $a\ne0$.}
\label{eta-vev}
$$
In one chirality this element only distinguishes even and odd operators, and may be omitted. Nevertheless, we retain it, since it will be necessary for a proper pairing of the two chiralities. The fermion energy\-/momentum tensor then reads
\eq$$
T_{bc}(z)=-{1\over2}\lcolon(\d\chi)^2\rcolon+{\i\over2}\d^2\chi.
\label{Tbc-chi}
$$
The currents read
\subeq{\label{J-chi}
\Align$$
J^N(z)
&=-\i\,\d\chi-\i\,\d X+{\i\over2}\,\d Y,
\label{JN-chi}
\\
J^E(z)
&=-\i\,\d Y,
\label{JE-chi}
\\
J^+(z)
&=\e^{-\gamma-\i\chi},
\label{J+-chi}
\\
J^-(z)
&=\i\lcolon(\d\chi-\d Y)\e^{\gamma+\i\chi}\rcolon,
\label{J--chi}
$$
}
Now define the vertex operators
\eq$$
\Phi^l_{en}(z)=\e^{l\gamma}V^l_{e\tn}(z)=\lcolon\e^{l\gamma+\i l\chi(z)-\i eX(z)-\i\tn Y(z)}\rcolon.
\label{Phil-def}
$$
Evidently, $\Phi^0_{en}=\Phi^+_{en}$, $\Phi^1_{en}=\Phi^-_{en}$. The conformal dimensions of the new vertex operators are given by
\eq$$
\Delta^l_{en}=\Delta_{en}+{l(l-1)\over2}.
\label{Delta-len-def}
$$
The number $l$ plays the role of a fermion number for the $bc$ system. Each vertex operator $\Phi^l_{en}$ corresponds to a vector $|v^l_{en}\rangle=\Phi^l_{en}(0)|\vac\rangle$, and free bosons $\chi$, $X$, $Y$ generate  a Fock module $\cF^l_{en}$ from this vector. The action of the charges on these vectors is given by
\eq$$
J^E_0|v^l_{en}\rangle=e|v^l_{en}\rangle,
\qquad
J^N_0|v^l_{en}\rangle=\left(n+{1\over2}-l\right)|v^l_{en}\rangle.
\label{JEN-vlen}
$$

On the other side, the elements $J^E_k$, $J^N_k$, $L_k$ and $\kappa$ form an extension of the Virasoro algebra, which we will denote by $\Vir^{EN}$ with the nonvanishing commutation relations
\eq$$
[J^E_k,J^N_{k'}]=k\kappa\delta_{k+k',0},
\quad
[L_k,L_{k'}]=(k-k')L_{k+k'},
\quad
[L_k,J^E_{k'}]=-k'J^E_{k+k'},
\quad
[L_k,J^N_{k'}]=-k'J^N_{k+k'}.
\label{VirEN-def}
$$
The action of this algebra on the highest weight vector $|v^l_{en}\rangle$ generates the irreducible module $\cV^l_{en}$. The Fock module $\cF^l_{en}$, the irreducible module $\cV^l_{en}$ and the Verma module (that is the module freely generated from the highest weight vector by the elements $L_{-k}$, $J^E_{-k}$, $J^N_{-k}$ with $k>0$) $M^l_{en}$ coincide as vector spaces for generic values of parameters $e,n$. Evidently, the spaces $\cV^l_{en}$ are eigenspaces of the operator $J^N_0$ with the eigenvalues $n+{1\over2}-l$. For generic values of $e,n$ the spaces $\cV^l_{en}$ are Verma modules of~$\Vir^{EN}$. But for $\ve\equiv e+l\in\Z$ the three types of modules are nonequivalent and the irreducible module $\cV^l_{en}$ is the factor of the Verma module $M^l_{en}$ over the level $|\ve-1/2|+1/2$ null vector of the Verma module. For $\ve\in\Z_{>0}$ we have $\cF^l_{en}\cong M^l_{en}$. For $\ve\in\Z_{\le0}$ the Fock module $\cF^l_{en}$ has the structure conjugate to a Verma module, and the irreducible module is a submodule of the Fock space generated from the highest weight vector. Below we will assume that the scalar product on these modules is normalized so that
\eq$$
\langle v^{l'}_{e'n'}|v^l_{en}\rangle=\delta_{e'e}\delta_{n'n}\delta_{l'l}.
\label{vlen-norm}
$$

Up to now nothing fixed the admissible values of the parameter $l$. In terms of the representation theory of the $\Vir^{EN}$ algebra it may take any real values. But the action of the operators $J^\pm_r$ with integer $r$ is only defined on the space $\cV^\text{NS}_{en}=\bigoplus_{l\in\Z}\cV^l_{en}$, so that is a $\widehat{gl}(1|1)_\text{NS}$ module. We will speak about these modules as of the Neveu\--Schwarz sector of the theory. On the contrary the operators $J^\pm_r$ with $r\in\Z+{1\over2}$ act on the space $\cV^{\text{R}}_{en}=\bigoplus_{l\in\Z+{1\over2}}\cV^l_{en}$, which is the highest weight module of the algebra $\widehat{gl}(1|1)_\text{R}$. Indeed, we have
\eq$$
J^{E,N}_k|v^{1/2}_{en}\rangle=0\quad(k\in\Z_{>0}),\qquad J^\pm_r|v^{1/2}_{en}\rangle=0\quad(r\in\Z_{>0}-{\textstyle{1\over2}}),
\label{R-hwv-def}
$$
so that $|v^{1/2}_{en}\rangle$ is a highest weight vector in the Ramond sector. It corresponds to the OPEs:
\subeq{\label{Phihalf-OPE}
\Align$$
J^\pm(z')\Phi^{1/2}_{en}(z)
&=O\left((z'-z)^{1/2}\right),
\label{Jpm-Phihalf-OPE}
\\
J^N(z')\Phi^{1/2}_{en}(z)
&={n\Phi^{1/2}_{en}(z)\over z'-z}+O(1),
\label{JN-Phihalf-OPE}
\\
J^E(z')\Phi^{1/2}_{en}(z)
&={e\Phi^{1/2}_{en}(z)\over z'-z}+O(1).
\label{JE-Phihalf-OPE}
$$
}

\begin{figure}[t]
\begin{subfigure}{.47\textwidth}
\center
\begin{tikzpicture}[scale=1.3]
\draw[fill=lightgray!50,thick] ([shift=(-92:.8)]-2,-3) -- (-2,-3) node[above left=-2pt] {$\cV^{-2}_{en}$} -- ([shift=(-88:.8)]-2,-3);
\draw[fill=lightgray!50,thick] ([shift=(-92:2.8)]-1,-1) -- (-1,-1) node[above left=-2pt] {$\cV^{-1}_{en}$} -- ([shift=(-88:2.8)]-1,-1);
\draw[fill=lightgray!50,thick] (-92:3.8) -- (0,0) node[above] {$\cV^0_{en}$} -- (-88:3.8);
\draw[fill=lightgray!50,thick] ([shift=(-92:3.8)]1,0) -- (1,0) node[above] {$\cV^1_{en}$} -- ([shift=(-88:3.8)]1,0);
\draw[fill=lightgray!50,thick] ([shift=(-92:2.8)]2,-1) -- (2,-1) node[above right=-2pt] {$\cV^2_{en}$} -- ([shift=(-88:2.8)]2,-1);
\draw[fill=lightgray!50,thick] ([shift=(-92:.8)]3,-3) -- (3,-3) node[above right=-2pt] {$\cV^3_{en}$} -- ([shift=(-88:.8)]3,-3);
\draw[thin] (-.2,0) -- (1.2,0);
\draw[thin] (-1.2,-1) -- (2.2,-1);
\draw[thin] (-1.2,-2) -- (2.2,-2);
\draw[thin] (-2.2,-3) -- (3.2,-3);
\draw[->] (0,0) -- node[above left=-3pt] {$J^+_{-1}$} (-1,-1);
\draw[->] (-1,-1) -- node[above left=-2pt] {$J^+_{-2}$} (-2,-3);
\draw[-] (-2,-3) -- (-2.267,-3.8);
\draw[->] (0,0) -- node[below=-2pt] {$J^-_0$} (1,0);
\draw[->] (1,0) -- node[above right=-3pt] {$J^-_{-1}$} (2,-1);
\draw[->] (2,-1) -- node[above right=-2pt] {$J^-_{-2}$} (3,-3);
\draw[-] (3,-3) -- (3.267,-3.8);
\end{tikzpicture}
\caption{Neveu\--Schwarz sector ($\cV^\text{NS}_{en}$)}
\end{subfigure}
\hfill
\begin{subfigure}{.47\textwidth}
\begin{tikzpicture}[scale=1.3]
\draw[fill=lightgray!50,thick] ([shift=(-92:1.8)]-2,-2) -- (-2,-2) node[above left=-2pt,xshift=2pt] {$\cV^{-{3\over2}}_{en}$} -- ([shift=(-88:1.8)]-2,-2);
\draw[fill=lightgray!50,thick] ([shift=(-92:3.3)]-1,-.5) -- (-1,-.5) node[above,xshift=-3pt] {$\cV^{-{1\over2}}_{en}$} -- ([shift=(-88:3.3)]-1,-.5);
\draw[fill=lightgray!50,thick] ([shift=(-92:3.8)]0,0) -- (0,0) node[above] {$\cV^{1\over2}_{en}$} -- ([shift=(-88:3.8)]0,0);
\draw[fill=lightgray!50,thick] ([shift=(-92:3.3)]1,-.5) -- (1,-.5) node[above,xshift=3pt] {$\cV^{3\over2}_{en}$} -- ([shift=(-88:3.3)]1,-.5);
\draw[fill=lightgray!50,thick] ([shift=(-92:1.8)]2,-2) -- (2,-2) node[above right=-2pt,xshift=-1pt] {$\cV^{5\over2}_{en}$} -- ([shift=(-88:1.8)]2,-2);
\draw[thin] (-.2,0) -- (.2,0);
\draw[thin] (-.2,-1) -- (.2,-1);
\draw[thin] (-2.2,-2) -- (-1.2,-2); \draw[thin] (-.8,-2) -- (.8,-2); \draw[thin] (1.2,-2) -- (2.2,-2);
\draw[thin] (-2.2,-3) -- (-1.2,-3); \draw[thin] (-.8,-3) -- (.8,-3); \draw[thin] (1.2,-3) -- (2.2,-3);
\draw[thin] (-1.2,-.5) -- (-.7,-.5); \draw[thin] (.7,-.5) -- (1.2,-.5);
\draw[thin] (-1.2,-1.5) -- (-.2,-1.5); \draw[thin] (.2,-1.5) -- (1.2,-1.5);
\draw[thin] (-1.2,-2.5) -- (-.2,-2.5); \draw[thin] (.2,-2.5) -- (1.2,-2.5);
\draw[thin] (-1.2,-3.5) -- (-.2,-3.5); \draw[thin] (.2,-3.5) -- (1.2,-3.5);
\draw[->] (0,0) -- node[below=-2pt,xshift=4pt] {$J^+_{-{1\over2}}$} (-1,-.5);
\draw[->] (-1,-.5) -- node[above left=-3pt] {$J^+_{-{3\over2}}$} (-2,-2);
\draw[-] (-2,-2) -- node[pos=.7,above left=-3pt] {$J^+_{-{5\over2}}$} (-2.72,-3.8);
\draw[->] (0,0) -- node[below=-2pt,xshift=-3pt] {$J^-_{-{1\over2}}$} (1,-.5);
\draw[->] (1,-.5) -- node[above right=-3pt,xshift=-3pt] {$J^-_{-{3\over2}}$} (2,-2);
\draw[-] (2,-2) -- node[pos=.7,above right=-3pt,xshift=-2pt] {$J^-_{-{5\over2}}$} (2.72,-3.8);
\end{tikzpicture}
\caption{Ramond sector ($\cV^\text{R}_{en}$)}
\end{subfigure}
\caption{The structure of the modules $\cV^\text{NS}_{en}$ and $\cV^\text{R}_{en}$. The currents that create the extremal vectors from the highest weight one are depicted by arrows. Horizontal lines show levels, where vectors exist in submodules.}
\label{fig-cVstruct}
\end{figure}
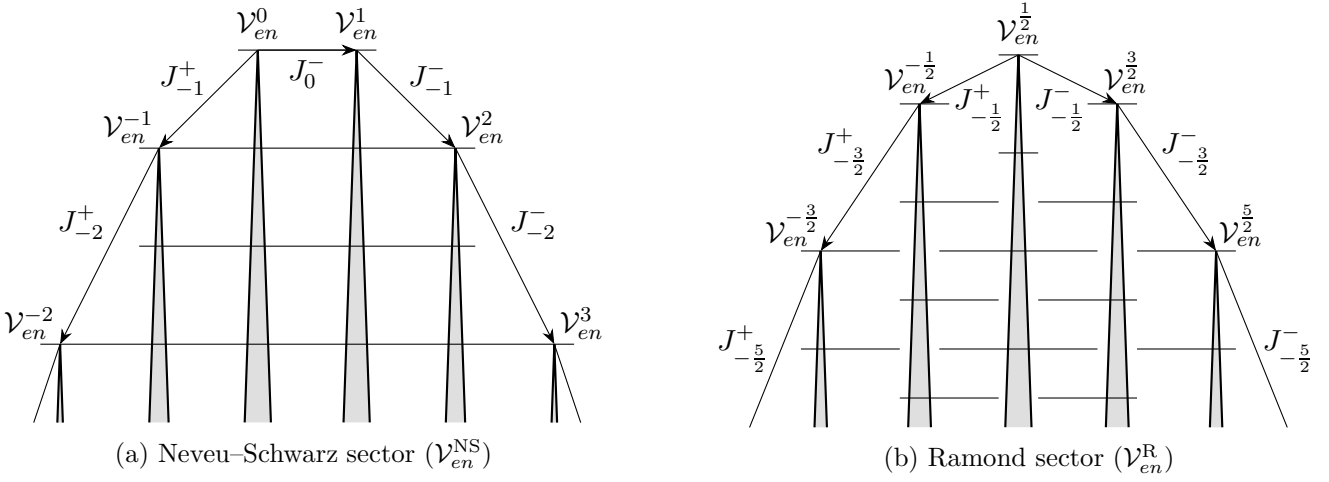

The more direct algebraic meaning of the vectors $|v^l_{en}\rangle$ is that they are extremal vectors in the representations~$\cV^\text{NS}_{en}$ and $\cV^\text{R}_{en}$ of the algebras $\widehat{gl}(1|1)_{\text{NS,R}}$ (see Fig.~\ref{fig-cVstruct}). Indeed, we have
\eq$$
\Aligned{
J^+_r|v^l_{en}\rangle
&=0\text{ for $r\ge l$},
&J^+_{l-1}|v^l_{en}\rangle
&=|v^{l-1}_{en}\rangle,
\\
J^-_r|v^l_{en}\rangle
&=0\text{ for $r\ge1-l$},
&J^-_{-l}|v^l_{en}\rangle
&=(e+l)|v^{l+1}_{en}\rangle.
}\label{vlen-extremal}
$$
It is assumed that $r-l\in\Z$.

For later use define a basis in the Verma module $M^l_{en}$. Let $K=(K^{(L)},K^{(E)},K^{(N)})$ a triple of partitions of integers (or Young diagrams):
$$
K^{(i)}\ (i=L,E,N):\quad k^{(i)}_1\ge k^{(i)}_2\ge\ldots\ge k^{(i)}_m\ge\ldots\ge0,
\qquad
\sum_nk^{(i)}_m=|K^{(i)}|<\infty.
$$
We also denote $|K|=|K^{(L)}|+|K^{(E)}|+|K^{(N)}|$. Define the operators
\eq$$
\Aligned{
\Lambda_{-K}
&=\prod^\curvearrowright_{m:\ k^{(L)}_m>0}L_{-k^{(L)}_m}\prod^\curvearrowright_{m:\ k^{(E)}_m>0}J^E_{-k^{(E)}_m}
\prod^\curvearrowright_{m:\ k^{(N)}_m>0}J^N_{-k^{(N)}_m},
\\
\Lambda_K
&=\prod^\curvearrowright_{m:\ k^{(N)}_m>0}J^N_{k^{(N)}_m}
\prod^\curvearrowright_{m:\ k^{(E)}_m>0}J^E_{k^{(E)}_m}\prod^\curvearrowright_{m:\ k^{(L)}_m>0}L_{k^{(L)}_m}.
}\label{Lambda-op-def}
$$
Then the vectors $\Lambda_{-K}|v^l_{en}\rangle$ form a basis in the module $\cV^l_{en}$, while the vectors $\langle v^l_{en}|\Lambda_K$ form a basis in the conjugate module.

%%%%%%%%%%%%%%%%%%%%%%%%%%%%%%%%%%%%%%%%%%%%%%%%%%%%%%%%%%%%%%%%%%%%%%%%
\subsection{Screening operator and fusion rules}

The screening operator $\cQ:\cV^l_{en}\to\cV^{l-1}_{e,n-1}$, which commutes with all the currents $J^\alpha(z)$ and, hence, with the energy\-/momentum tensor is
\eq$$
\cQ(C)=\oint_C dz\,P(z),
\qquad
P(z)=b(z)V_{0,-1}(z)=\e^{-\gamma}V^{-1}_{0,-1}(z).
\label{Screening-def}
$$
The screening operator depends on a closed contour $C$ on the covering surface of the complex plane. This screening operator is of the fermion type:
\eq$$
(\cQ(C))^2=0.
\label{Q2=0}
$$
It means that each field $\Phi^l_{en}(z)$ has two quantum group components.

To obtain fusion rules, we have to study the products $\Phi^{l_1}_{e_1n_2}(z)\Phi^{l_2}_{e_2n_2}(0)$. For the sake of brevity we will often write
$$
\Phi_i(z)\equiv\Phi^{l_i}_{e_in_i}(z),
\qquad
V_i(z)\equiv V^{l_i}_{e_i\tn_i}(z).
$$
Without screening operators they are obtained by a straightforward multiplication:
\Multline*$$
\Phi_1(z)\Phi_2(0)= z^{\delta_{12}}\e^{(l_1+l_2)\gamma}\lcolon V_1(z)V_2(0)\rcolon
=z^{\delta_{12}}\e^{(l_1+l_2)\gamma}\left(V^{l_1+l_2}_{e_1+e_2,\tn_1+\tn_2}(z)+O(z)\right)
\\
=z^{\delta_{12}}\left(\Phi^{l_1+l_2}_{e_1+e_2,n_1+n_2+{1\over2}}(0)+O(z)\right).
$$
Here
\eq$$
\delta_{ij}=e_i\tn_j+e_j\tn_i+l_il_j.
\label{deltaij-def}
$$

Now consider products with one screening. For $\ve_i=e_i+l_i\not\in\Z$ we may consider the closed `eyeglasses rim' contour \cite{Felder:1988zp} or, for simplicity, its contraction to a line from $0$ to $z$ with a standard regularization. We have
\Multline*$$
\Phi_1(z)\cQ^z_0\Phi_2(0)
=z^{\delta_{12}}\e^{(l_1+l_2-1)\gamma}\int^z_0dw\,(z-w)^{-\ve_1}w^{-\ve_2}
\lcolon V^{l_1}_{e_1\tn_1}(z)V^{-1}_{0,-1}(w)V^{l_2}_{e_2\tn_2}(0)\rcolon
\\
=z^{\delta_{12}}\e^{(l_1+l_2-1)\gamma}\int^z_0dw\,(z-w)^{-\ve_1}w^{-\ve_2}
\left(V^{l_1+l_2-1}_{e_1+e_2,\tn_1+\tn_2-1}(0)+O(z)\right)
\\
=z^{\delta_{12}-\ve_1-\ve_2+1}\left(\beta_{12}\Phi^{l_1+l_2-1}_{e_1+e_2,n_1+n_2-{1\over2}}(0)+O(z)\right),
$$
where $\cQ^{z'}_z=\int^{z'}_zdw\,P(w)$ and
$$
\beta_{ij}={\Gamma(1-\ve_i)\Gamma(1-\ve_j)\over\Gamma(2-\ve_i-\ve_j)}.
$$
Finally, the operator product expansions read
\eq$$
\Aligned{
\Phi^{l_1}_{e_1n_1}(z)\Phi^{l_2}_{e_2n_2}(0)
&=z^{\delta_{12}}\left(\Phi^{l_1+l_2}_{e_1+e_2,n_1+n_2+{1\over2}}(0)+O(z)\right),
\\
\Phi^{l_1}_{e_1n_1}(z)\cQ^z_0\Phi^{l_2}_{e_2n_2}(0)
&=z^{\delta_{12}-\ve_1-\ve_2+1}
\left(\beta_{12}\Phi^{l_1+l_2-1}_{e_1+e_2,n_1+n_2-{1\over2}}(0)+O(z)\right).
}\label{PhiPhi-OPE}
$$
The two lines of this expressions are, in a sense, `symmetric' to each other. This symmetry originates in the automorphism of the algebra $\widehat{gl}(1|1)$
\eq$$
\Gathered{
J^+\rightarrow J^-,
\qquad
J^-\rightarrow-J^+
\qquad
J^{N,E}\to-J^{N,E},
\\
\Phi^l_{en}\to q_l(e)\Phi^{1-l}_{-e,-n},
}
\label{JV-automorph}
$$
where
\eq$$
\Aligned{
q_l(e)
&=(-e)^{-1/2}\prod^{l-1}_{k=1}(-e-k)^{-1},
&&\text{if $l\in\Z$,}
\\
q_l(e)
&=\prod^{l-1/2}_{k=1}\left(\half-e-k\right)^{-1},
&&\text{if $l\in\Z+\half$,}
}\label{qle-def}
$$
or, more uniformly,
\eq$$
q_l(e)=(-e)^{\{l\}-1\over2}{\Gamma(1-e-l)\over\Gamma(\{l\}-e)}
\label{qle-uniform}
$$
with $\{l\}$ being the fractional part of~$l$. Evidently, $\{l\}=0$ for an NS operator and $\{l\}={1\over2}$ for an R one.

In deriving these functions we fixed the values $q_1(e)=(-e)^{-1/2}$ and $q_{1/2}(e)=1$. The first one implies the symmetry between the $(-e)^{-1/4}\Phi^+_{en}$ and $(-e)^{1/4}\Phi^-_{en}$ operators, while the second one means that the $\Phi^{1/2}_{en}$ maps to itself. The generality is not lost due to the automorphism $J^\pm_r\to\Lambda^{\pm1}J^\pm_r$ ($\Lambda\in\C$).

The automorphism (\ref{JV-automorph}) provides an alternative representation
\eq$$
\Phi^l_{en}=\e^{(1-l)\gamma}V^{1-l}_{-e,1-\tn},
\label{V-alternative}
$$
which yields expressions equivalent to those obtained from (\ref{J-chi}), (\ref{Phil-def}) subject to $\ve=e+l\not\in\Z$.

The expansions (\ref{PhiPhi-OPE}) correspond to the well known fusion rule for typical fields of $GL(1|1)$ WZW model with $e+e'\neq 0$:
\eq$$
[\Phi_{e'n'}][\Phi_{en}]\sim[\Phi_{e'+e,n'+n+{1\over2}}]+[\Phi_{e'+e,n'+n-{1\over2}}].
\label{PhiPhi-fusion}
$$

\subsection{Two-point chiral correlation functions}
\label{sec-two-point-chiral}

Consider two\-/point correlation functions in the (right) chiral sector. Since
$$
\langle[J^\alpha_0,Z]\rangle
=\langle\vac|J^\alpha_0Z|\vac\rangle-\langle\vac|ZJ^\alpha_0|\vac\rangle=0,
\qquad\alpha=N,E,
$$
where $Z$ is any operator, we have
\eq$$
\langle\Phi^{l_1}_{e_1n_1}(z_1)\Phi^{l_2}_{e_2n_2}(z_2)\rangle
=(z_1-z_2)^{-2\Delta^{l_1}_{e_1n_1}}\delta_{e_1+e_2,0}\delta_{n_1+n_2,0}\delta_{l_1+l_2,1},
\label{2p-chiral}
$$
where $\delta_{a,b}$ should be understood as the Kronecker symbol rather than as the delta function though its arguments are continuous. In the free field realization it means that
\eq$$
\langle V^{l_1}_{e_1\tn_1}(z')V^{l_2}_{e_2\tn_2}(z)\rangle
=(z'-z)^{-2\Delta^{l_1}_{e_1n_1}}\delta_{e_1+e_2,0}\delta_{\tn_1+\tn_2,1}\delta_{l_1+l_2,1}.
\label{VcV-2p-chiral}
$$
Hence, the expectation value $\langle\cdots\rangle$ is defined with `charge at infinity', i.e.
\eq$$
\langle Z\rangle=\lim_{\zeta\to\infty}\zeta\langle P(\zeta)Z\rangle_{\text{free field}},
\label{expval-def}
$$
where $Z$ is any operator realized in terms of free fields. This is consistent with the requirement of the conformal invariance of the $XY$ and $bc$ systems.

%%%%%%%%%%%%%%%%%%%%%%%%%%%%%%%%%%%%%%%%%%%%%%%%%%%%%%%%%%%%%%%%%%%%%%%%
\subsection{Two chiralities}
\label{sec-two-chiralities}

Equations (\ref{J-def}) only define the currents in the right chiral sector. For the left chirality we may introduce the free fields $\bX(\bz),\allowbreak\bY(\bz),\allowbreak\bb(\bz),\allowbreak\bc(\bz)$ with the same properties and set
\subeq{\label{bJ-def}
\Align$$
\bJ^N(\bz)
&=\lcolon\bb\bc\rcolon-\i\,\d\bX+{\i\over2}\,\bd\bY,
\label{bJN-def}
\\
\bJ^E(\bz)
&=-\i\,\bd\bY,
\label{bJE-def}
\\
\bJ^+(\bz)
&=\bb,
\label{bJ+-def}
\\
\bJ^-(\bz)
&=-\i c\,\bd\bY+\bd\bc.
\label{bJ--def}
$$
}
We assume odd currents of different chiralities anticommuting:
$$
J^\alpha_r\bJ^\beta_{r'}=-\bJ^\beta_{r'}J^\alpha_r,
\qquad
\alpha,\beta=\pm.
$$
Correspondingly the fermion operators $b(z),c(z)$ anticommute with $\bb(z),\bc(z)$. This assumption simplifies formulas. Analogously to (\ref{bc-chi}) we introduce the field $\bar\chi(\bz)$ and the algebraic element $\bgamma$:
\eq$$
\bb(z)=\lcolon\e^{-\bgamma-\i\bar\chi(\bz)}\rcolon,
\qquad
\bc(z)=\lcolon\e^{\gamma+\i\bar\chi(\bz)}\rcolon,
\label{bc-chi-bar}
$$
The anticommutativity of right and left chiral fermion operators imposes the commutation relations
\eq$$
[\gamma,\bgamma]=-\i\pi.
\label{gamma-bargamma-alg}
$$
This results in
\eq$$
\e^{a\gamma}\e^{b\bgamma}=\e^{-\i\pi ab}\e^{b\bgamma}\e^{a\gamma}.
\label{gamma-bargamma-exp-commut}
$$

The left chiral vertex operators read
\eq$$
\bar\Phi^\bl_{en}(\bz)=\e^{\bl\bgamma}\bar V^l_{e\tn}(\bz)
=\e^{\bl\bgamma+\i\bl\bar\chi(\bz)-e\bX(\bz)-\tn\bY(\bz)}\rcolon.
\label{Phil-bar-def}
$$
Then the local operator, which corresponds to the vertex operators $\Phi^l_{en}(z)$, $\bar\Phi^\bl_{en}(\bz)$, has the form%
\footnote{Though formally this definition differs from that of \cite{Schomerus:2005bf}, it is equivalent due to the automorphism~(\ref{JV-automorph}).}
\eq$$
\Aligned{
\Phi^{l\bl}_{en}(z,\bz)
&=\e^{l\gamma}\e^{\bl\bgamma}V^{l\bl}_{e\tn}(z,\bz)
=\e^{l\gamma}\e^{\bl\bgamma}V^l_{e\tn}(z)V^\bl_{e\tn}(\bz).
}\label{Phi-tot-def}
$$
We will assume that both $l,\bl\in\Z$ in the NS sector and both $l,\bl\in\Z+{1\over2}$ in the R sector. These operators have the spin
\eq$$
\sigma^{l\bl}=\Delta^l_{en}-\Delta^\bl_{en}={l(l-1)\over2}-{\bl(\bl-1)\over2}.
\label{Phi-spin}
$$
In the NS sector the spin is always integer, while in the R sector it is integer if $l-\bl$ is even, and half\-/integer if $l-\bl$ is odd.

It is important that, due to the anticommutation of $\bJ^\pm_r$ and $\e^{\pm\gamma}$ the action of $\bJ^\pm_r$ takes an additional factor $\e^{\pm\i\pi l}$, so that the relation (\ref{vlen-extremal}) is modified
\eq$$
\Aligned{
J^+_{l-1}|\Phi^{l\bl}_{en}\rangle
&=|\Phi^{l-1,\bl}_{en}\rangle,
&\bJ^+_{\bl-1}|\Phi^{l\bl}_{en}\rangle
&=\e^{\i\pi l}|\Phi^{l,\bl-1}_{en}\rangle,
\\
J^-_{-l}|\Phi^{l\bl}_{en}\rangle
&=(e+l)|\Phi^{l+1,\bl}_{en}\rangle,
&\bJ^-_{-\bl}|\Phi^{l\bl}_{en}\rangle
&=\e^{-\i\pi l}(e+\bl)|\Phi^{l,\bl+1}_{en}\rangle,
}\label{Phi-full-extremal}
$$
The screening operator is
\eq$$
\cQ=\int d^2z\,P(z,\bz),
\qquad
P(z,\bz)=\e^{-\bgamma}\e^{-\gamma}V^{-1,-1}_{0,-1,-1}(z,\bz).
\label{Scr-tot-def}
$$

The automorphism (\ref{JV-automorph}) in the right chiral sector reads
\eq$$
\Gathered{
\bJ^+\rightarrow\bJ^-,
\qquad
\bJ^-\rightarrow-\bJ^+
\qquad
\bJ^{N,E}\to-\bJ^{N,E},
\\
\bar\Phi^\bl_{en}(\bz)\to q_\bl(e)\bar\Phi^{1-\bl}_{-e,-n}(\bz).
}
\label{barJV-automorph}
$$
Thus there is an alternative representation
\eq$$
\bar\Phi^\bl_{en}(\bz)=\e^{(1-\bl)\bgamma}\bar V^{1-\bl}_{-e,1-\tn}(\bz).
\label{barV-alternative}
$$
For the operators $\Phi^{l\bl}_{en}(z,\bz)$ of the field theory this automorphism produces the following symmetry:
\eq$$
\Phi^{l\bl}_{en}(z,\bz)
\to(-1)^{\lceil\bl\rceil-1}Q_{l\bl}(e)\Phi^{1-l,1-\bl}_{-e,-n}(z,\bz),
\label{pm-symm}
$$
where the sign $\lceil\cdot\rceil$ designates the ceiling of a number and
\eq$$
Q_{l\bl}(e)=q_l(e)q_\bl(e)={\Gamma(1-e-l)\Gamma(1-e-\bl)\over\Gamma(\{l\}-e)\Gamma(1-\{l\}-e)}.
\label{Qlle-def}
$$
Below we impose the symmetry (\ref{pm-symm}) on all correlation functions.

%%%%%%%%%%%%%%%%%%%%%%%%%%%%%%%%%%%%%%%%%%%%%%%%%%%%%%%%%%%%%%%%%%%%%%%%
\section{Correlation functions}
\label{sec-correlation-functions}

In this section we study correlation functions of the form
\eq$$
G_{12\ldots N}(x_1,\ldots,x_N)
=\langle\Phi_N(x_N)\cdots\Phi_2(x_2)\Phi_1(x_1)\rangle,
\qquad
\Phi_i(x)=\Phi^{l_i\bl_i}_{e_in_i}(z,\bz).
\label{G-gen}
$$
The average sign $\langle\cdots\rangle$ means the vacuum\-/vacuum matrix element. We should normalize the fields properly. Assume that
\eq$$
\Gathered{
\langle\Phi_2(x_2)\Phi_1(x_1)\rangle
=\Gamma_{12}(z_2-z_1)^{-2\Delta_1}(\bz_2-\bz_1)^{-2\bar\Delta_1},
\\*
\Gamma_{ij}=
\Gamma_i\delta_{e_i+e_j,0}\delta_{n_i+n_j,0}\delta_{l_i+l_j,1}\delta_{\bl_i+\bl_j,1},
\qquad
\Gamma_i=\e^{\i\pi l_i(1-\bl_i)},
}\label{PhiPhi-corr}
$$
where $\Delta_i=\Delta^{l_i}_{e_in_i}$, $\bar\Delta_i=\Delta^{\bl_i}_{e_in_i}$. Assuming
\eq$$
\langle\Phi_2|\Phi_1\rangle=\delta_{12}\equiv\delta_{e_1e_2}\delta_{n_1n_2}\delta_{l_1l_2}\delta_{\bl_1\bl_2},
\label{PhiPhi-matel}
$$
we obtain
\eq$$
\langle\Phi^{l\bl}_{en}|=\langle\vac|\Phi^{1-l,1-\bl}_{-e,-n}(\infty,\infty)\e^{-\i\pi l(1-\bl)},
\label{Phi-conj-def}
$$
where
\eq$$
\Phi^{l\bl}_{en}(\infty,\infty)
=\lim_{\zeta\to\infty}\zeta^{2\Delta^l_{en}}\bar\zeta^{2\Delta^\bl_{en}}
\Phi^{l\bl}_{en}(\zeta,\bar\zeta)
\label{Phi-inf-def}
$$

The factor $\Gamma_{ij}$ takes into account the commutation relation
\eq$$
\Phi_i(z',\bz')\Phi_j(z,\bz)=Z_{ij}\Phi_j(z,\bz)\Phi_i(z,\bz'),
\qquad
Z_{ij}=(-1)^{(l_i-\bl_i)(l_j+\bl_j)}.
\label{PhiPhi-commut}
$$
There is an important point here. The mutual locality demands that $Z_{ij}=Z_{ji}$, but for two Ramond operators this condition is only satisfied if $l_j-\bl_j=l_i-\bl_i\bmod2$. Hence, the Ramond sector is split into two subsectors of mutually local operators:
\eq$$
\Aligned{
\text{\Rp}:\quad
&l-\bl\in2\Z,
\\
\text{\Rm}:\quad
&l-\bl\in2\Z+1.
}\label{Rpm-sectors}
$$
We recognize in these subsectors the spaces of operators of integer and half\-/integer spins. Operators in each of the Ramond subsectors are mutually local with those in the NS sector, but operators from different subsectors are mutually semilocal.

Correlation functions of the conformal field theory can be expressed in terms of the free field correlation functions as follows:
\eq$$
G_{12\ldots N}(x_1,\ldots,x_N)
=\cN_{12\ldots N}\lim_{\zeta\to\infty}|\zeta|^2
\langle P(\zeta,\bar\zeta)\Phi_N(x_N)\ldots\Phi_2(x_2)\Phi_1(x_1)\cQ^{L-1}\rangle_{\text{free field}}.
\label{G-ff}
$$
Here $\cN_{12\ldots N}$ is a normalization factor (equal to 1 in the case $N=2$, $L=1$) and
\eq$$
L=\sum^N_{i=1}l_i=\sum^N_{i=1}\bl_i.
\label{s-def}
$$
Due to the symmetry (\ref{pm-symm}) there is a bound
\eq$$
1\le L\le N-1
\label{s-upperbound}
$$
for nonzero correlation functions. Due to the $J^E_0$ and $J^N_0$ invariance of the vacuum, we have two more restrictions
\eq$$
\sum^N_{i=1}e_i=0,
\qquad
\sum^N_{i=1}n_i={N\over2}-L.
\label{en-zero}
$$
Our aim is to calculate the three\-/point functions, which has the standard form
\eq$$
G_{123}(x_1,x_2,x_3)
=C_{123}\prod^3_{\substack{i<j\\k\ne i,j}}(z_j-z_i)^{\Delta_k-\Delta_i-\Delta_j}
(\bz_j-\bz_i)^{\bar\Delta_k-\bar\Delta_i-\bar\Delta_j}.
\label{G123}
$$
Here $\Delta_i=\Delta^{l_i}_{e_in_i}$, $\bar\Delta_i=\bar\Delta^{\bl_i}_{e_in_i}$ and the coefficients $C_{123}$ are defined in terms of the structure constants of the operator algebra $C_{12}{}^3$:
\eq$$
C_{123}=\sum_{3'}C_{12}{}^{3'}\Gamma_{3'3},
\label{C123-strconst}
$$
where the summation is taken over the variables $l'_3,\bl'_3,e'_3,n'_3$ denoted here as $3'$. Due to the antidiagonal form of $\Gamma$ the only admissible value is $l'_3=l_3^*$, $\bl'_3=\bl_3^*$, $e'_3=e_3^*$, $n'_3=n_3^*$, where the star means the conjugation
\eq$$
l^*=1-l,\qquad\bl^*=1-\bl,\qquad e^*=-e,\qquad n^*=-n.
\label{len-conj}
$$

It is well known~\cite{Belavin:1984vu} that the structure constants cannot be found without studying four\-/point correlation functions, which we proceed with. All four\-/point correlation functions can be obtained from the functions
\eq$$
G_{1234}(z,\bz)
=\langle\Phi_4(\infty,\infty)\Phi_3(1,1)\Phi_2(z,\bz)\Phi_1(0,0)\rangle
\label{G1234-def}
$$
by a Möbius transformation. Besides, in what follows we assume that $\ve_i\not\in\Z$, $\ve_i+\ve_j\not\in\Z$. According to~\cite{Belavin:1984vu} these functions admit a decomposition
\eq$$
G_{1234}(z,\bz)
=\sum_{s,4'}\Gamma_{4'4}C_{12}{}^sC_{s3}{}^{4'}
\bF_s\sbm{2&3\\1&4'}(z)\bbF_s\sbm{2&3\\1&4'}(\bz),
\label{G1234-decomp}
$$
where the conformal blocks $\bF_s\sbm{2&3\\1&4}(z)$ are labeled by the `right mover' triples: $i\to(l_i,e_i,n_i)$. To avoid multiple bars in our notation we use the symbol $\bbF$ for the complex conjugate functions, which are labeled by the `left mover' triples: there $i\to(\bl_i,e_i,n_i)$. The label $s$ corresponds to the intermediate conformal family in the $s$\-/channel. Here we assume the `canonical' normalization of conformal blocks:
\eq$$
\bF_s\sbm{2&3\\1&4}(z)=z^{\Delta_s-\Delta_1-\Delta_2}(1+O(z)).
\label{bF-norm}
$$
In what follows we will also need some other normalizations.

The condition that the intermediate state $s$ is a physical one leads to the fact that the values of $l_s$ and $\bl_s$ are uniquely defined by the value of $n_s$, which according to the fusion rule (\ref{PhiPhi-fusion}) admits no more than two values:
\eq$$
n_s=n_1+n_2+{1\over2}-\alpha=-n_3-n_4-{1\over2}+\beta,
\qquad
\alpha,\beta=0,1.
\label{ns-values}
$$
For given values of $\alpha,\beta$ we have
\eq$$
l_s=l_1+l_2-\alpha=1-l_3-l_4+\beta,
\qquad
\bl_s=\bl_1+\bl_2-\alpha=1-\bl_3-\bl_4+\beta.
\label{ls-values}
$$
This relation establishes a one-to-one correspondence between the triples denoted by $s$ and $\bar s$ in the r.h.s.\ of~(\ref{G1234-decomp}), which makes sense to the summation over~$s$.

For the future use, let us introduce the notation $s(\alpha),\ldots,u'(\alpha)$:
\eq$$
\Aligned{
n_{s(\alpha)}
&=n_1+n_2+{1\over2}-\alpha,\quad
&n_{s'(\alpha)}
&=-n_3-n_4-{1\over2}+\alpha,
\\
n_{t(\alpha)}
&=n_3+n_2+{1\over2}-\alpha,
&n_{t'(\alpha)}
&=-n_1-n_4-{1\over2}+\alpha,
\\
n_{u(\alpha)}
&=n_1+n_3+{1\over2}-\alpha,
&n_{u'(\alpha)}
&=-n_2-n_4-{1\over2}+\alpha.
}\label{stu(alpha)-def}
$$

We will say that the decomposition (\ref{G1234-decomp}) is the $s$\-/channel one. The structure constants are determined from the crossing symmetry~\cite{Belavin:1984vu}, which is the identity involving the $s$- and $t$\-/channel decompositions:
\eq$$
\sum_sC_{12}{}^sC_{s3}{}^{4^*}
\bF_s\sbm{2&3\\1&4^*}(z)\bbF_s\sbm{2&3\\1&4^*}(\bz)
=Z_{123}\sum_tC_{32}{}^tC_{t1}{}^{4^*}
\bF_t\sbm{2&1\\3&4^*}(1-z)\bbF_t\sbm{2&1\\3&4^*}(1-\bz).
\label{crossing}
$$
Here we introduced a sign factor
\eq$$
Z_{123}=Z_{12}Z_{13}Z_{23}\e^{-\i\pi\sum^4_{i=1}\sigma_i}
=\e^{\i\pi\left(\sum^3_{i\ne j}(l_j\bl_i-l_i\bl_j)+(L-l_4-\bl_4)(l_4-\bl_4)\right)}.
\label{Z123-def}
$$
This factor comes from the fact that the sum in the l.h.s.\ corresponds to the correlation function of the form $\langle\Phi_4\Phi_3\Phi_2\Phi_1\rangle$, while that in the r.h.s.\ corresponds to the correlation function of the form $\langle\Phi_4\Phi_1\Phi_2\Phi_3\rangle$. It is equal to $\pm1$, if all four operators $\Phi_i$ are mutually local.

We postpone a systematic treatment of crossing and braiding properties to section~\ref{sec-braiding-crossing}. Now we perform the conformal block decompositions in terms of the free field theory, and establish the structure constants.

From (\ref{ns-values}) immediately follows that in a four\-/point correlation function $\sum_i n_i$ can take one of the three values $-1,0,1$, which correspond to $L=1,2,3$. In the first and third cases there is only one intermediate channel $p$, which corresponds to $\alpha=\beta=0$ or $1$ respectively. In the second case there are two intermediate channels with $\alpha=0$, $\beta=1$ and with $\alpha=1$, $\beta=0$. We will consider all three cases.

{\bf1.}~$L=1$. In this case $\sum n_i=-1$ and there is only one intermediate family in each channel:
\eq$$
s=s(0)=s'(0),
\qquad
t=t(0)=t'(0),
\qquad
u=u(0)=u'(0).
\label{stu-L1}
$$
In the $t$\-/channel the situation is the same up to the substitution $n_1\leftrightarrow n_3$. In the free field construction this corresponds to a correlation function without screening operators
\Align$$
G_{1234}(z,\bz)
&\sim\langle V_4(\infty,\infty)V_3(1,1)V_2(z,\bz)
V_1(0,0)\rangle
\notag
\\
&=z^{\delta_{12}}(1-z)^{\delta_{23}}\bz^{\bar\delta_{12}}(1-\bz)^{\bar\delta_{23}}
\equiv H_{123}(z,\bz).
\label{G-L1}
$$
Here we denoted
\eq$$
\bar\delta_{ij}=e_i\tn_j+e_j\tn_i+\bl_i\bl_j.
\label{bardeltaij-def}
$$
The notation $H_{123}$ is introduced for the future use.

By representing (\ref{G-L1}) in the form (\ref{G1234-decomp}) and taking into account the antidiagonal form of~$\Gamma$, we obtain
\Align$$
G_{1234}(z,\bz)
&=\Gamma_{4^*}C_{12}{}^sC_{s3}{}^{4^*}
z^{\delta_{12}}(1-z)^{\delta_{23}}\bz^{\vphantom{1^1}\bar\delta_{12}}(1-\bz)^{\bar\delta_{23}}
\notag
\\
&=\Gamma_s^{-1}C_{12s^*}C_{s34}\,
z^{\delta_{12}}(1-z)^{\delta_{23}}\bz^{\vphantom{1^1}\bar\delta_{12}}(1-\bz)^{\bar\delta_{23}},
\label{G-L1-CBexpansion}
$$
We have
\eq$$
\bF_s\sbm{2&3\\1&4^*}(z)=z^{\delta_{12}}(1-z)^{\delta_{23}}=\bF_t\sbm{2&1\\3&4^*}(1-z).
\label{cF-L1}
$$
Then, in terms of the constants $C_{ijk}$ the crossing symmetry condition reads
\eq$$
\Gamma_s^{-1}C_{12s^*}C_{s34}=Z_{123}\Gamma_t^{-1}C_{32t^*}C_{t14},
\label{crossing-L1}
$$
where $i^*\sim(1-l_i,-e_i,-n_i)$.

{\bf2.}~$L=2$. In this case $\sum n_i=0$ and we have two intermediate states in each channel:
\eq$$
s=s(\alpha)=s'(1-\alpha),
\qquad
t=t(\alpha)=t'(1-\alpha),
\qquad
u=u(\alpha)=u'(1-\alpha),
\qquad
\alpha=0,1.
\label{stu-L2}
$$
In the free field representation such correlation functions contain one screening operator. First, let us calculate
\Align$$
G_{1234}(z,\bz)
&\sim\langle V_4(\infty,\infty)V_3(1,1)V_2(z,\bz)V_1(0,0)\cQ\rangle
\notag
\\
&=H_{123}(z,\bz)
\int d^2w\,w^{-\ve_1}\bw^{-\bve_1}(w-z)^{-\ve_2}(\bw-\bz)^{-\bve_2}(w-1)^{-\ve_3}(\bw-1)^{-\ve_3},
\label{G-L2-int}
$$
where again
\eq$$
\ve_i=e_i+l_i,
\qquad
\bve_i=e_i+\bl_i.
\label{ve-def}
$$
Evidently, for the case $L=2$ we have
\eq$$
\ve_1+\ve_2+\ve_3+\ve_4=\bve_1+\bve_2+\bve_3+\bve_4=2.
\label{sumve2}
$$
In the $s$\-/channel decomposition this correlation function reads
\eq$$
{G_{1234}(z,\bz)\over H_{123}(z,\bz)}
=\sum^1_{\alpha=0}
X_\alpha I_\alpha(\ve_1,\ve_2,\ve_3;z)I_\alpha(\bve_1,\bve_2,\bve_3;\bz),
\label{G-L2-s-decomp}
$$
where $X_\alpha$ are constants and $I_\alpha$ are the Dotsenko\--Fateev integrals~\cite{Dotsenko:1984nm}:
\eq$$
\Aligned{
I_0(a,b,c;z)
&=\int^\infty_1dw\,w^{-a}(w-z)^{-b}(w-1)^{-c}
\\
I_1(a,b,c;z)
&=\int^z_0dw\,w^{-a}(z-w)^{-b}(1-w)^{-c},
}\label{Ialpha-def}
$$
with the usual regularization: the integrals are understood as an analytic continuation from the convergence region in the space of parameters $a,b,c$. Define the conformal blocks in the Dotsenko\--Fateev normalization:%
\footnote{Later, in Sec~\ref{sec-braiding-crossing} we will discuss different normalizations of conformal blocks. There we will use special notation for each of them. In this section the letter $\cF$ will always denote conformal blocks in the Dotsenko\--Fateev normalization.}
\eq$$
\cF_{s(\alpha)}\sbm{2&3\\1&4^*}(z)=z^{\delta_{12}}(1-z)^{\delta_{23}}I_\alpha(\ve_1,\ve_2,\ve_3;z).
\label{cFDF-L1-def}
$$
By comparing (\ref{G-L2-s-decomp}) with the decomposition~(\ref{G1234-decomp}), we conclude that
\eq$$
\cF_s\sbm{2&3\\1&4^*}(z)=K_s\sbm{2&3\\1&4^*}\bF_s\sbm{2&3\\1&4^*}(z),
\label{cFDF-L1-cF}
$$
where $K_s\sbm{2&3\\1&4^*}$ is a normalization factor. It is known that in the limit $|z|\ll1$
\eq$$
\Aligned{
I_0(a,b,c;z)
&\simeq{\Gamma(1-c)\Gamma(a+b+c-1)\over\Gamma(a+b)},
\\
I_1(a,b,c;z)
&\simeq z^{1-a-b}{\Gamma(1-a)\Gamma(1-b)\over\Gamma(2-a-b)}.
}\label{Ialpha-lim}
$$
Hence,
\eq$$
K_{s(\alpha)}\sbm{2&3\\1&4^*}=K_\alpha(\alpha_1,\alpha_2,\alpha_2)\equiv\Cases{
  {\Gamma(1-\ve_3)\Gamma(1-\ve_4)\over\Gamma(2-\ve_3-\ve_4)},
  &\text{if $\alpha=0$;}
  \\
  {\Gamma(1-\ve_1)\Gamma(1-\ve_2)\over\Gamma(2-\ve_1-\ve_2)},
  &\text{if $\alpha=1$.}
}
\label{Ks-fin}
$$
These normalization factors relate the constants $X_\alpha=X_{s(\alpha)}\sbm{2&3\\1&4^*}$ with products of structure constants:
\eq$$
X_s\sbm{2&3\\1&4^*}K_s\sbm{2&3\\1&4^*}\bar K_s\sbm{2&3\\1&4^*}
=\Gamma_s^{-1}C_{12s^*}C_{s34},
\label{X-CC-rel}
$$
where $\bar K_{s(\alpha)}\sbm{2&3\\1&4^*}=K_\alpha(\bve_1,\bve_2,\bve_3)$.

Similarly, in the $t$\-/channel we have
\eq$$
{G_{1234}(z,\bz)\over H_{123}(z,\bz)}
=\sum^1_{\alpha=0}
X'_\alpha I_\alpha(\ve_3,\ve_2,\ve_1;1-z)I_\alpha(\bve_3,\bve_2,\bve_1;1-\bz),
\label{G-L2-t-decomp}
$$
where $X'_\alpha=X_{t(\alpha)}\sbm{2&1\\3&4^*}$ are related to structure constants according to
\eq$$
Z_{123}X_t\sbm{2&1\\3&4^*}K_t\sbm{2&1\\3&4^*}\bar K_t\sbm{2&1\\3&4^*}
=\Gamma_t^{-1}C_{32t^*}C_{t14}.
\label{X'-CC-rel}
$$

The integrals $I_\alpha(z)$ and $I_\alpha(1-z)$ are related as follows~\cite{Dotsenko:1984nm,Dotsenko:1984ad}:
\eq$$
I_\alpha(a,b,c;z)=\sum^1_{\beta=0}F_{\alpha\beta}(a,b,c)I_\beta(c,b,a;1-z)
\label{Ialpha-trans}
$$
with the matrix of coefficients
\eq$$
F(a,b,c)={1\over\sin\pi(b+c)}\pMatrix{\sin\pi a&-\sin\pi b\\-\sin\pi(a+b+c)&-\sin\pi c}.
\label{fij-coeffs}
$$
By substituting (\ref{Ialpha-trans}) into (\ref{G-L2-s-decomp}) and comparing it with (\ref{G-L2-t-decomp}) we obtain
\eq$$
\sum^1_{\alpha=0}X_\alpha F_{\alpha\beta}(\ve_1,\ve_2,\ve_3)F_{\alpha\gamma}(\bve_1,\bve_2,\bve_3)=X'_\beta\delta_{\beta\gamma}.
\label{crossing-XX'}
$$

By taking $\beta\ne\gamma$ we easily obtain
\eq$$
{X_0\over X_1}=-{\sin\pi\bve_3\sin\pi\ve_4\over\sin\pi\ve_1\sin\pi\bve_2}.
\label{X1X0-ratio}
$$
By substituting this into (\ref{crossing-XX'}) for $\beta=\gamma=0$, we obtain
\eq$$
{X_0\over X'_0}={\sin\pi\bve_3\sin\pi(\ve_3+\ve_2)\over\sin\pi\ve_1\sin\pi(\bve_1+\bve_2)}.
\label{X0X'0-ratio}
$$
From these equations and (\ref{X-CC-rel}), (\ref{X'-CC-rel}) we immediately find
\Align$$
{C_{12s(0)^*}C_{s(0)34}
\over C_{12s(1)^*}C_{s(1)34}}
&=-{\Gamma_{s(0)}\over\Gamma_{s(1)}}
{\Gamma(\ve_1)\Gamma(\bve_2)\Gamma(1-\ve_3)\Gamma(1-\bve_4)\Gamma(2-\ve_1-\ve_2)\Gamma(2-\bve_1-\bve_2)
  \over\Gamma(1-\bve_1)\Gamma(1-\ve_2)\Gamma(\bve_3)\Gamma(\ve_4)\Gamma(\ve_1+\ve_2)\Gamma(\bve_1+\bve_2)},
\label{CCCC-s0s1-ratio}
\\
{C_{12s(0)^*}C_{s(0)34}\over C_{32t(0)^*}C_{t(0)14}}
&={\Gamma_{s(0)}\over\Gamma_{t(0)}}
{\Gamma(\ve_1)\Gamma(1-\ve_3)\Gamma(1-\bve_1-\bve_2)\Gamma(\ve_2+\ve_3)
  \over\Gamma(1-\bve_1)\Gamma(\bve_3)\Gamma(\ve_1+\ve_2)\Gamma(1-\ve_2-\ve_3)}.
\label{CCCC-s0t0-ratio}
$$
Note that neither these ratios nor the symmetry properties~(\ref{pm-symm}) contain any $n_i$. We will search the structure constants in an $n$\-/independent form. To establish them consider two types of correlation functions.

{\bf2a.}~4NS correlation functions: $l_i\in\Z$ ($i=1,\ldots,4$). In this case $l_s,l_t\in\Z$. Hence, by studying these four\-/point functions, we compute the structure constants for three NS operators.

The symmetry (\ref{pm-symm}) leads to the equation
\eq$$
C_{123}=(-)^{\sum\lceil\bl_i\rceil-1}C_{1^*2^*3^*}\prod^3_{i-1}Q_i,
\label{C123-symm}
$$
where $Q_i=Q_{l_i\bl_i}(e_i)$ in terms of the functions defined in~(\ref{Qlle-def}). By applying it to (\ref{CCCC-s0s1-ratio}) we obtain
\eq$$
\left(C_{12s(0)^*}\over C_{1^*2^*s(1)}\right)\left(C_{s(1)34}\over C_{s(0)^*3^*4^*}\right)^{-1}
=(-1)^{l_1-\bl_1+l_3-\bl_3}{R_{12}\over R_{34}},
\label{CCCC-s0s1-*-ratio}
$$
where
\eq$$
R_{12}={\Gamma(1-\{l_1\}+e_1)\Gamma(1-\{l_2\}+e_2)\Gamma(1-\{l_1+l_2\}-e_1-e_2)
  \over\Gamma(1-\{l_1\}-e_1)\Gamma(1-\{l_2\}-e_2)\Gamma(1-\{l_1+l_2\}+e_1+e_2)}.
\label{R-def}
$$
The fractional parts $\{l_i\}=l_i-\lfloor l_i\rfloor$ are introduced for the later use in the $\rm R_\pm$ sectors. Let
\eq$$
\e^{-\i\pi(l_1\bl_2+(l_1+l_2)(1-\bl_1-\bl_2))}C_{12s(0)^*}
=C(e_1,l_1,\bl_1;e_2,l_2,\bl_2)
\equiv C_{12}.
\label{C12-def}
$$
Then all structure constants in~(\ref{CCCC-s0s1-*-ratio}) are expressed in terms of the function~$C_{ij}$:
\eq$$
\left(C_{12}\over C_{1^*2^*}\right)\left(C_{34}\over C_{3^*4^*}\right)^{-1}
={R_{12}\over R_{34}}.
\label{CCCC-s0s1-sqrt-ratio}
$$
For the special case $3=1^*$, $4=2^*$ we obtain
\eq$$
{C_{12}\over C_{1^*2^*}}=\pm R_{12}.
\label{CC*-ratio}
$$

Turn now to the equation (\ref{CCCC-s0t0-ratio}). By applying (\ref{C123-symm}) to the second factors in the ratio of the l.h.s., we obtain
\Align$$
{C_{12s(0)^*}C_{s(0)^*3^*4^*}\over C_{32t(0)^*}C_{t(0)^*1^*4^*}}
&=(-1)^{(1-l_1-l_3)(\bl_1+\bl_3)}
\notag
\\*
&\quad\times
{\Gamma(1+e_1)\Gamma(1-e_3)\Gamma(1+e_2+e_3)\Gamma(1-e_1-e_2)
  \over\Gamma(1-e_1)\Gamma(1+e_3)\Gamma(1-e_2-e_3)\Gamma(1+e_1+e_2)}.
\label{CCCC-*-X0X'0-NS-ratio}
$$
In terms of the functions $C_{ij}$ it takes the form
\eq$$
{C_{12}C_{3^*4^*}\over C_{32}C_{1^*4^*}}
=\left(R_{12}R_{14}\over R_{32}R_{34}\right)^{1/2}.
\label{CCCC-*-X0X'0-red-ratio}
$$
This equation is much more restrictive than (\ref{CCCC-s0s1-sqrt-ratio}) and has the only (up to an overall sign, which is the same for all values of $l_i,\bl_i$) solution
\eq$$
C_{12}=R_{12}^{1/2}.
\label{C(0)-final}
$$
It is consistent with equation (\ref{CC*-ratio}) for the `$+$' sign and with equation~(\ref{crossing-L1}).

{\bf2b.}~2R2NS correlation functions: $l_1,l_2\in\Z+{1\over2}$, $l_3,l_4\in\Z$. In this case we have $l_s\in\Z$, $l_t\in\Z+{1\over2}$. Thus one can compute the structure constants for two R operators and one NS operator.

In this case, by applying (\ref{C123-symm}) to (\ref{CCCC-s0s1-ratio}) we obtain
\eq$$
\left(C_{12s(0)^*}\over C_{1^*2^*s(1)}\right)\left(C_{s(1)34}\over C_{s(0)^*3^*4^*}\right)^{-1}
=(-1)^{l_2-\bl_1+l_4-\bl_3+1}{R_{12}\over R_{34}}.
\label{CCCC-s0s1-*-R-ratio}
$$
Hence,
\eq$$
{C_{12s(0)^*}\over C_{1^*2^*s(1)}}=(-1)^{l_1-\bl_2}R_{12}.
\label{CC-R-ratio}
$$
It is easy to check that the only solution consistent with the commutation relations for two R operators and one NS operator and the normalization (\ref{PhiPhi-corr}) is again (\ref{C(0)-final}) with the definition (\ref{C12-def}).

Now return to arbitrary $l_i\in{1\over2}\Z$. We see that the formula (\ref{C(0)-final}) is correct in the case $\sum l_i=1$. In the case $\sum l_i=2$ we may apply (\ref{C123-symm}) and use the identity
$$
(-)^{\lceil\bl\rceil-1}Q_{l\bl}(e){\Gamma(1-\{l\}-e)\over\Gamma(1-\{l\}+e)}={\Gamma(1-\ve)\over\Gamma(\bar\ve)}.
$$
Finally, we obtain
\eq$$
C_{123}=
\Gamma_{123}R_{12}^{1/2}\times\Cases{1,&\text{if $\sum l_i=1$;}\\
 (-)^{l_2-\bl_2}S_{12},&\text{if $\sum l_i=2$,}}
\label{C123-final}
$$
where
\eq$$
\Gamma_{12\ldots n}=\e^{\i\pi\sum^n_{i<j}l_i\bl_j}.
\label{Gamma12...n-def}
$$
and
\eq$$
S_{12}={\Gamma(1-\ve_1)\Gamma(1-\ve_2)\Gamma(\ve_1+\ve_2-1)\over\Gamma(\bar\ve_1)\Gamma(\bar\ve_2)\Gamma(2-\bar\ve_1-\bar\ve_2)}.
\label{S12-def}
$$
Note that the signs of these structure constants are only well\-/defined, if all three operators are in the NS sector or if there is one NS operator and two R operators that belong to the same sector either \Rp\ or \Rm.

For practical calculations of four\-/point functions the values of $X_\alpha$ are also useful:
\eq$$
\Aligned{
X_0
&={\sin\pi\ve_3\sin\pi\bve_4\over\pi\sin\pi(\ve_3+\ve_4)}X_{1234},
\\
X_1
&={\sin\pi\bve_1\sin\pi\ve_2\over\pi\sin\pi(\ve_1+\ve_2)}X_{1234},
}\label{Xi-final}
$$
where
\eq$$
X_{1234}=\Gamma_{1234}\left(\prod^4_{i=1}{\Gamma(1-\{l_i\}+e_i)\over\Gamma(1-\{l_i\}-e_i)}\right)^{1/2}.
\label{X1234-def}
$$

{\bf3.}~$L=3$. In this case $\sum n_i=1$ and we have only one intermediate state:
\eq$$
s=s(1)=s'(1),
\qquad
t=t(1)=t'(1),
\qquad
u=u(1)=u'(1).
\label{stu-L3}
$$
Surely, correlation functions for $L=3$ are obtained from those in the case $L=1$ by means of the substitution~(\ref{pm-symm}). Nevertheless, it is instructive to obtain them from the free field representation. In short, correlation functions in this sector contain two screening operators, e.g.
\eq$$
G_{1234}(z,\bz)
\sim\langle V_4(\infty,\infty)V_3(1,1)V_2(z,\bz)V_1(0,0)\cQ^2\rangle.
\label{G-L2}
$$
Since $(\cQ(C))^2=0$, the two screenings must go along two different contours in each chiral sector. Hence
$$
G_{1234}(z,\bz)
=H_{123}(z,\bz)I^{(2)}(\ve_1,\ve_2,\ve_3;z)I^{(2)}(\bve_1,\bve_2,\bve_3;\bz),
$$
where
\Multline*$$
I^{(2)}(a,b,c;z)=\int^\infty_1dw_1\int^z_0dw_2\,w_1^{-a}(w_1-z)^{-b}(w_1-1)^{-c}w_2^{-a}(z-w_2)^{-b}(1-w_2)^{-c}(w_1-w_2)
\\
=I_1(a-1,b,c;z)I_2(a,b,c;z)-I_1(a,b,c;z)I_2(a-1,b,c;z).
$$
Its convergence region is $a,b,c>-1$, $a+b+c<-2$. For $|z|\ll1$ we have
$$
I^{(2)}(a,b,c;z)\sim z^{1-a-b}.
$$
On the other hand
$$
I^{(2)}(a,b,c;z)\sim I^{(2)}(c,b,a;1-z)
$$
and for $|1-z|\ll1$ we have
$$
I^{(2)}(a,b,c;z)\sim(1-z)^{1-c-b}.
$$
Besides, by manipulating with integrals it is not difficult to derive that
$$
I^{(2)}(z)=O(z^{2-a-2b-c})\quad\text{as $z\to\infty$}
$$
in the convergence region. It means that
$$
I^{(2)}(a,b,c;z)\sim z^{1-a-b}(1-z)^{1-c-b}F(z)
$$
with $F(z)$ being a bounded integer functions, which is nothing but a constant. After a simple calculation we obtain
\eq$$
I^{(2)}(a,b,c;z)
=-{\sin\pi(a+b)\over\pi}\Gamma(1-a)\Gamma(1-b)\Gamma(1-c)z^{1-a-b}(1-z)^{1-c-b}.
\label{I(2)-final}
$$
Hence,
\eq$$
G_{1234}(z,\bz)=\Gamma_s^{-1}C_{12s^*}C_{s34}H_{1^*2^*3^*}(z,\bz)
\label{G-L2-fin}
$$
in consistency with (\ref{G-L1}) and~(\ref{pm-symm}).

%%%%%%%%%%%%%%%%%%%%%%%%%%%%%%%%%%%%%%%%%%%%%%%%%%%%%%%%%%%%%%%%%%%%%%%%
\section{Braiding, crossing and Moore\--Seiberg equations}
\label{sec-braiding-crossing}

Correlation functions of a conformal field theory are constructed in terms of local, i.e.\ commuting or anticommuting, operators, so they admit different conformal block decompositions. This imposes a set of equations on conformal blocks called braiding and crossing properties. Below we will need conformal blocks in different normalizations. Thus we will consider $\cF_s\sbm{2&3\\1&4}(z)=K_s\sbm{2&3\\1&4}\bF_s\sbm{2&3\\1&4}(z)$ with arbitrary constant factors $K_s\sbm{2&3\\1&4}$. The braiding and crossing equations have the form
\Align$$
z^{\Delta_4-\sum^3_{i=1}\Delta_i}\cF_s\sbm{2&3\\1&4}(z^{-1})
&=\sum_uB_{su}\sbm{2&3\\1&4}^\pm\cF_u\sbm{3&2\\1&4}(z),
\label{bm}
\\
\cF_s\sbm{2&3\\1&4}(1-z)
&=\sum_tF_{st}\sbm{2&3\\1&4}\cF_t\sbm{2&1\\3&4}(z),
\label{fm}
$$
where the coefficients $B\sbm{2&3\\1&4}^\pm(s,u)$ and $F\sbm{2&3\\1&4}(s,t)$ are called braiding and crossing matrices respectively. The superscript $\pm$ denotes the two possible ways for analytic continuation in the complex plane from $z$ to $z^{-1}$ above and below $1$ correspondingly.

Let us represent these equations graphically. Omitting details related to the $z$ variable, denote
$$
\cF_s\sbm{2&3\\1&4}(z)
=\vcenter{\hbox{
\begin{tikzpicture}
\draw[->] (-1.5,0) -- (-1,0) node[anchor=north east] {$1$};
  \draw[->] (-1,0) -- (0,0) node[anchor=north] {$s$};
  \draw[->] (0,0) -- (1,0) node[anchor=north west] {$4$}; \draw[-] (1,0) -- (1.5,0);
\draw[->] (-.5,-1) -- (-.5,-.5) node[anchor=north east] {$2$}; \draw[-] (-.5,-.5) -- (-.5,0);
\draw[->] (.5,-1) -- (.5,-.5) node[anchor=north west] {$3$}; \draw[-] (.5,-.5) -- (.5,0);
\end{tikzpicture}
}}
$$
Thus, equations (\ref{bm}), (\ref{fm}) can be written as follows
\Align*$$
\vcenter{\hbox{
\begin{tikzpicture}
\draw[->] (-1.5,0) -- (-1,0) node[anchor=north east] {$1$};
  \draw[->] (-1,0) -- (0,0) node[anchor=north] {$s$};
  \draw[->] (0,0) -- (1,0) node[anchor=north west] {$4$}; \draw[-] (1,0) -- (1.5,0);
\draw[->] (-.5,-1) -- (-.5,-.5) node[anchor=north east] {$2$}; \draw[-] (-.5,-.5) -- (-.5,0);
\draw[->] (.5,-1) -- (.5,-.5) node[anchor=north west] {$3$}; \draw[-] (.5,-.5) -- (.5,0);
\end{tikzpicture}
}}
&=\sum_uB_{su}\sbm{2&3\\1&4}^+
\vcenter{\hbox{
\begin{tikzpicture}
\draw[->] (-1.5,0) -- (-1,0) node[anchor=north east] {$1$};
  \draw[->] (-1,0) -- (0,0) node[anchor=north] {$u$};
  \draw[->] (0,0) -- (1,0) node[anchor=north west] {$4$}; \draw[-] (1,0) -- (1.5,0);
\draw[->] (.5,-1.5) node[anchor=south west] {$3$} to[out=90,in=-90] (-.5,-.3); \draw[-] (-.5,-.3) -- (-.5,0);
\fill[white] (0,-.9) circle(.2);
\draw[->] (-.5,-1.5) node[anchor=south east] {$2$} to[out=90,in=-90] (.5,-.3); \draw[-] (.5,-.3) -- (.5,0);
\end{tikzpicture}
}}
\\[\medskipamount]
\vcenter{\hbox{
\begin{tikzpicture}
\draw[->] (-1.5,0) -- (-1,0) node[anchor=north east] {$1$};
  \draw[->] (-1,0) -- (0,0) node[anchor=north] {$s$};
  \draw[->] (0,0) -- (1,0) node[anchor=north west] {$4$}; \draw[-] (1,0) -- (1.5,0);
\draw[->] (-.5,-1) -- (-.5,-.5) node[anchor=north east] {$2$}; \draw[-] (-.5,-.5) -- (-.5,0);
\draw[->] (.5,-1) -- (.5,-.5) node[anchor=north west] {$3$}; \draw[-] (.5,-.5) -- (.5,0);
\end{tikzpicture}
}}
&=\sum_tF_{st}\sbm{2&3\\1&4}
\vcenter{\hbox{
\begin{tikzpicture}
\draw[->] (-1,0) -- (-.5,0) node[anchor=north east] {$1$};
  \draw[->] (-.5,0) -- (.5,0) node[anchor=north west] {$4$}; \draw[-] (.5,0) -- (1,0);
\draw[->] (0,-1) -- (0,-.4) node[anchor=west] {$t$}; \draw (0,-.4) -- (0,0);
\draw[->] (-.5,-1.9) node[anchor=north] {$2$} -- (-.5,-1.6); \draw[-] (-.5,-1.6) to[out=90,in=180] (0,-1);
\draw[->] (.5,-1.9) node[anchor=north] {$3$} -- (.5,-1.6); \draw[-] (.5,-1.6) to[out=90,in=0] (0,-1);
\end{tikzpicture}
}}
$$

In the fundamental work by Moore and Seiberg~\cite{Moore:1988qv} important equations for braiding and crossing matrices, called hexagon and pentagon equations, were found as consistency conditions. It was shown that for some particular sets of the normalization factors $K_s\sbm{2&3\\1&4}$ these equations have a rather simple form. The hexagon equation reads
\eq$$
\sum_pB_{6p}\sbm{2&3\\1&7}^\pm B_{79}\sbm{2&4\\p&5}^\pm B_{p8}\sbm{3&4\\1&9}^\pm
=\sum_pB_{7p}\sbm{3&4\\6&5}^\pm B_{68}\sbm{2&4\\1&p}^\pm B_{p9}\sbm{2&3\\8&5}^\pm,
\label{hex}
$$
while the pentagon equations read
\Align$$
F_{68}\sbm{2&3\\1&7} F_{79}\sbm{8&4\\1&5}
&=\sum_pF_{7p}\sbm{3&4\\6&5}F_{69}\sbm{2&p\\1&5}F_{p8}\sbm{3&2\\4&9},
\label{penFF}
\\
F_{69}\sbm{2&3\\1&7}B_{78}\sbm{9&4\\1&5}^\pm
&=\sum_pB_{7p}\sbm{3&4\\6&5}^\pm B_{68}\sbm{2&4\\1&p}^\pm F_{p9}\sbm{2&3\\8&5}.
\label{penFB}
$$

There are a few simpler equations, which also can be derived from the basic principles of rational CFT:
\Align$$
\sum_pB_{5p}\sbm{2&3\\1&4}^\pm B_{p6}\sbm{3&2\\1&4}^\mp
&=\delta_{56},
\label{BBinv}
\\
\sum_pF_{5p}\sbm{2&3\\1&4}F_{p6}\sbm{2&1\\3&4}
&=\delta_{56},
\label{FFinv}
\\
\sum_pB_{5p}\sbm{2&3\\1&4}^\pm F_{p6}\sbm{3&2\\1&4}
&=F_{56}\sbm{2&3\\1&4}\,\e^{\mp\i\pi(\Delta_2+\Delta_3-\Delta_6)},
\label{BF}
\\
F_{56}\sbm{2&3\\1&4}
&=B_{56}\sbm{1&3\\2&4}^\pm\,\e^{\mp\i\pi(\Delta_2+\Delta_4-\Delta_5-\Delta_6)}.
\label{BeqF}
$$

In what follows we derive the correct normalization factors $K_s\sbm{2&3\\1&4}$, which provide this simple form of the Moore\--Seiberg equations in the case of $\widehat{gl}(1|1)$ model for $\ve_i\not\in\Z$.

In the normalization (\ref{bF-norm}) (i.e.\ for $K_s\sbm{2&3\\1&4}=1$) the braiding and crossing matrices do not satisfy the equations (\ref{hex})--(\ref{penFB}). We need another basis. Let us construct the appropriate basis starting from the integrals that appear in the free field representation. Recall that there are two natural bases depending on the form of contours. In Sect.\ \ref{sec-correlation-functions} we used the Dotsenko\--Fateev (DF) basis~\cite{Dotsenko:1984nm,Dotsenko:1984ad}, where contours connect points, where vertex operators are situated. Generally, for four\-/point conformal blocks it has the form
\eq$$
\begin{tikzpicture}
\draw (-.5,0) node[anchor=east] {$\cI^{DF}_{L,\alpha}(z)={}$};
\draw[fill] (0,0) circle(2pt) node[anchor=east] {$0$};
\draw[fill] (3.6,0) circle(2pt) node[anchor=west] {$z$};
\draw[fill] (6,0) circle(2pt) node[anchor=east] {$1$};
\draw[->] (0,0) to [out=15,in=180] (1.8,0.3) node[anchor=south] {$\alpha$};
  \draw[-] (1.8,.3) to[out=0,in=165] (3.6,0);
\draw (1.8,0) node {$\strut\vdots$};
\draw[->] (0,0) to [out=-15,in=180] (1.8,-.3); \draw[-] (1.8,-.3) to[out=0,in=-165] (3.6,0);
\draw[->] (6,0) to [out=10,in=-175] (8,.3) node[anchor=south] {$L-1-\alpha$};
  \draw[-] (8,.3) to[out=5,in=-178] (8.5,.32);
\draw (8,0) node {$\strut\vdots$};
\draw[->] (6,0) to [out=-10,in=175] (8,-.3); \draw[-] (8,-.3) to[out=-5,in=178] (8.5,-.32);
\end{tikzpicture}
\label{cIDF-def}
$$
The total number of contours $L-1$ is constant, while the integer $\alpha=0,\ldots,L-1$ enumerates the elements of the basis. In our case, as we already said, $L=1,2,3$ and we have four integrals in total:
\eq$$
\Aligned{
\cI^{DF}_{1,0}(z)
&=z^{\delta_{12}}(1-z)^{\delta_{23}},
\\
\cI^{DF}_{2,0}(z)
&=z^{\delta_{12}}(1-z)^{\delta_{23}}I_0(-\ve_1,-\ve_2,-\ve_3;z),
\\
\cI^{DF}_{2,1}(z)
&=z^{\delta_{12}}(1-z)^{\delta_{23}}I_1(-\ve_1,-\ve_2,-\ve_3;z),
\\
\cI^{DF}_{3,1}(z)
&=z^{\delta_{12}}(1-z)^{\delta_{23}}I^{(2)}(-\ve_1,-\ve_2,-\ve_3;z),
}\label{cIDF-Ii}
$$
Thus the DF basic conformal blocks read
\eq$$
\cF^{DF}_s\sbm{2&3\\1&4}(z)
=\cI^{DF}_{L,\alpha}(z),
\qquad
l_1+l_2+l_3=L+l_4-1,
\qquad
l_s=l_1+l_2-\alpha.
\label{cFDF-def}
$$

Another basis, proposed by Felder~\cite{Felder:1988zp}, consists of loop contours around zero:
\eq$$
\begin{tikzpicture}
\draw (-2.9,0) node[anchor=east] {$\cI^{Fe}_{L,\alpha}(z)={}$};
\draw[fill] (0,0) circle(2pt) node[anchor=west] {$0$};
\draw[fill] (3.6,0) circle(2pt) node[anchor=west] {$z$};
\draw[fill] (6,0) circle(2pt) node[anchor=west] {$1$};
\draw[-] (3.6,0) to[out=170,in=0] (0,.5);
  \draw[->] (0,.5) arc (90:180:.5); \draw[-] (-.5,0) arc (180:270:.5);
  \draw[-] (0,-.5) to[out=0,in=-170] (3.6,0);
\draw (-.8,0) node {$\stackrel{\alpha}\cdots$};
\draw[-] (3.6,0) to[out=165,in=0] (0,.8);
  \draw[->] (0,.8) to[out=180,in=90] (-1.1,0); \draw[-] (-1.1,0) to[out=-90,in=180] (0,-.8);
  \draw[-] (0,-.8) to[out=0,in=-165] (3.6,0);
\draw[-] (6,0) to[out=170,in=0] (0,1.1);
  \draw[->] (0,1.1) to[out=180,in=90] (-1.5,0); \draw[-] (-1.5,0) to[out=-90,in=180] (0,-1.1);
  \draw[-] (0,-1.1) to[out=0,in=-170] (6,0);
\draw (-2.1,0) node {$\stackrel{L-1-\alpha}\cdots$};
\draw[-] (6,0) to[out=165,in=0] (0,1.4);
  \draw[->] (0,1.4) to[out=180,in=90] (-2.7,0); \draw[-] (-2.7,0) to[out=-90,in=180] (0,-1.4);
  \draw[-] (0,-1.4) to[out=0,in=-165] (6,0);
\end{tikzpicture}
\label{cIFe-def}
$$
An advantage of this basis is that it is easily and uniformly generalized to an arbitrary number of points. Moreover, it admits an easy factorization of a multipoint conformal block into smaller ones.

By contracting the contours (\ref{cIFe-def}) to $0$ and $\infty$, we obtain
\eq$$
\Aligned{
\cI^{Fe}_{1,0}(z)
&=\cI^{DF}_{1,0}(z),
\\
\cI^{Fe}_{2,0}(z)
&=2\i\sin\pi\ve_4\,\cI^{DF}_{2,0}(z),
\\
\cI^{Fe}_{2,1}(z)
&=-2\i\sin\pi\ve_1\,\cI^{DF}_{2,1}(z),
\\
\cI^{Fe}_{3,1}(z)
&=-4\sin\pi\ve_1\sin\pi\ve_4\,\cI^{DF}_{3,1}(z).
}\label{cI-Fe-DF-rel}
$$
Similarly, the Felder basic conformal blocks read
\eq$$
\cF^{Fe}_s\sbm{2&3\\1&4}(z)
=\cI^{Fe}_{L,\alpha}(z),
\qquad
l_1+l_2+l_3=L+l_4-1,
\qquad
l_s=l_1+l_2-\alpha.
\label{cFFe-def}
$$
Though these two bases only differ in the normalization, the effect for the form of the hexagon and pentagon equations turns out to be crucial. The $B$ matrix in the Felder basis satisfies the hexagon equation in the form~(\ref{hex}), while that in the DF basis does not. In the DF basis neither hexagon nor pentagon equation are satisfied. Nevertheless, below we construct a basis, in which all hexagon and pentagon equations (\ref{hex})--(\ref{penFB}) are satisfied simultaneously, and this basis can be considered as a modification of the DF one.

First of all, define matrices $B^\pm(a,b,c)$ by the equations
\eq$$
z^{1-a-b-c}I_\alpha(a,b,c;z^{-1})
=\sum^1_{\beta=0}B^\pm_{\alpha\beta}(a,b,c)I_\beta(a,c,b;z),
\label{Ii-braiding}
$$
where the superscript $\pm$ depends on the way of the analytic continuation from $z$ to $z^{-1}$ around~$1$. We easily obtain
\eq$$
B^\pm(a,b,c)
={1\over\sin\pi(a+c)}\pMatrix{-\e^{\mp\i\pi(a+b+c)}\sin\pi b&-\e^{\mp\i\pi b}\sin\pi a\\
  -\e^{\mp\i\pi c}\sin\pi(a+b+c)&\e^{\pm\i\pi a}\sin\pi c}.
\label{bij-coeffs}
$$
Together with the matrix $F(a,b,c)$ from (\ref{fij-coeffs}) it provides the braiding and crossing in the DF basis. For the $B$ matrix in the DF basis we have
\eq$$
B^{DF}_{\alpha\beta}\sbm{2&3\\1&4}^\pm
=\e^{\pm\i\pi\delta_{23}}\times\Cases{1,
  &\text{if $L=1$;}
  \\
  B^\pm_{\alpha\beta}(\ve_1,\ve_2,\ve_3),
  &\text{if $L=2$;}
  \\
  \e^{\mp\i\pi(\ve_2+\ve_3+1)}{\sin\pi(\ve_1+\ve_2)\over\sin\pi(\ve_1+\ve_3)},
  &\text{if $L=3$.}
}
\label{BDF-fin}
$$
Recall that for $L=1$ we have $\alpha=\beta=0$, while for $L=3$ we have $\alpha=\beta=1$. The $F$ matrix in the DF basis reads
\eq$$
F^{DF}_{\alpha\beta}\sbm{2&3\\1&4}
=\Cases{1,
  &\text{if $L=1$;}
  \\
  F_{\alpha\beta}(\ve_1,\ve_2,\ve_3),
  &\text{if $L=2$;}
  \\
  {\sin\pi(\ve_1+\ve_2)\over\sin\pi(\ve_3+\ve_2)},
  &\text{if $L=3$.}
}
\label{FDF-fin}
$$
In the Felder basis we easily obtain
\eq$$
B^{Fe}_{\alpha\beta}\sbm{2&3\\1&4}^\pm
=\e^{\pm\i\pi\delta_{23}}\times\Cases{1,
  &\text{if $L=1$;}
  \\
  \tilde B^\pm_{\alpha\beta}(\ve_1,\ve_2,\ve_3),
  &\text{if $L=2$;}
  \\
  \e^{\mp\i\pi(\ve_2+\ve_3+1)}{\sin\pi(\ve_1+\ve_2)\over\sin\pi(\ve_1+\ve_3)},
  &\text{if $L=3$,}
}
\label{BFe-fin}
$$
and
\eq$$
F^{Fe}_{\alpha\beta}\sbm{2&3\\1&4}
=\Cases{1,
  &\text{if $L=1$;}
  \\
  \tilde F_{\alpha\beta}(\ve_1,\ve_2,\ve_3),
  &\text{if $L=2$;}
  \\
  {\sin\pi\ve_1\sin\pi(\ve_1+\ve_2)\over\sin\pi\ve_3\sin\pi(\ve_3+\ve_2)},
  &\text{if $L=3$,}
}
\label{FFe-fin}
$$
where
\Align$$
\tilde B^\pm_{\alpha\beta}(a,b,c)
&={1\over\sin\pi(a+c)}\pMatrix{-\e^{\mp\i\pi(a+b+c)}\sin\pi b&-\e^{\mp\i\pi b}\sin\pi(a+b+c)\\
  -\e^{\mp\i\pi c}\sin\pi a&\e^{\pm\i\pi a}\sin\pi c},
\label{Babc-Fe-def}
\\
\tilde F_{\alpha\beta}(a,b,c)
&={\sin\pi a\over\sin\pi(b+c)}\pMatrix{1&-{\sin\pi(a+b+c)\sin\pi b\over\sin\pi a\sin\pi c}\\1&-1}.
\label{Fabc-Fe-def}
$$

It can be checked that the matrix $B^{Fe}\sbm{1&2\\3&4}(s,u)$ satisfies the hexagon equation (\ref{hex}). The reason is that the many\-/point basic conformal blocks in the Felder basis possess a good cluster factorization property. Nevertheless, the matrix $F^{Fe}\sbm{1&2\\3&4}(s,t)$ does not satisfy the pentagon equations. Due to {\ref{cI-Fe-DF-rel}} the DF and Felder bases only differ by normalization factors. Let us make the transformation from the Felder basis to DF one for the case $L=2$, but for $L=1,3$ normalization changes we impose the condition that the hexagon equation would remain valid. It turns out that in the resulting basis both the hexagon and the pentagon equations are obeyed.

Suppose, for some normalization of conformal blocks the $B$ matrices satisfy the hexagon equation~(\ref{hex}). Let $\cF'_s\sbm{2&3\\1&4}(z)=f_s\sbm{2&3\\1&4}\cF_s\sbm{2&3\\1&4}(z)$. Then the $B$ matrix $B'_{su}\sbm{2&3\\1&4}^\pm$ related to $\cF'$ are related to $B_{su}\sbm{2&3\\1&4}^\pm$ according to
$$
B_{su}\sbm{2&3\\1&4}^\pm=\left(f_s\sbm{2&3\\1&4}\right)^{-1}B'_{su}\sbm{2&3\\1&4}^\pm f_u\sbm{3&2\\1&4}.
$$
After substituting it into (\ref{hex}), we obtain that the matrices $B'$ satisfy the hexagon equation if the ratio
$$
g(p)={f_p\sbm{3&2\\1&7}f_9\sbm{4&2\\p&5}\over f_p\sbm{3&4\\1&9}f_7\sbm{2&4\\p&5}}
$$
is $p$ independent for admissible values of~$p$.

There are two admissible values of $p$ in $f_p\sbm{3&2\\1&4}$, if $L=2$. Hence, take the second and third lines of~(\ref{cI-Fe-DF-rel}) to define these function for the transformation from the Felder to the DF basis. For $L=2$ we have
$$
f_{s(\alpha)}\sbm{2&3\\1&4}=\Cases{2\i\sin\pi\ve_4,&\text{if $\alpha=0$;}\\
  -2\i\sin\pi\ve_1,&\text{if $\alpha=1$.}}
$$
Then the $p$ independence of $g(p)$ condition admits a solution
$$
f_s\sbm{2&3\\1&4}=\Cases{(-2\i\sin\pi(\ve_1+\ve_2))^{-1},&\text{if $L=1$;}\\
  8\i\sin\pi(\ve_1+\ve_2)\sin\pi\ve_1\sin\pi\ve_4,&\text{if $L=3$.}}
$$
This solution corresponds to the following basis
\eq$$
\cF^{MS}_s\sbm{2&3\\1&4}(z)
=\cF^{DF}_s\sbm{2&3\\1&4}(z)\times\Cases{-2\i\sin\pi(\ve_1+\ve_2),&\text{if $L=1$;}\\
  1,&\text{if $L=2$;}\\
  (-2\i\sin\pi(\ve_1+\ve_2))^{-1},&\text{if $L=3$.}}
\label{cFMS-def}
$$
We shall refer to this basis as to the Moore\--Seiberg (MS) basis, since the $B$ and $F$ matrices in this basis satisfy both hexagon and pentagon equations~(\ref{hex})--(\ref{penFB}). Explicitly, these matrices read
\Align$$
B^{MS}_{\alpha\beta}\sbm{2&3\\1&4}^\pm
&=\e^{\pm\i\pi\delta_{23}}\times\Cases{{\sin\pi(\ve_1+\ve_2)\over\sin\pi(\ve_1+\ve_3)},
  &\text{if $L=1$;}
  \\
  B^\pm_{\alpha\beta}(\ve_1,\ve_2,\ve_3),
  &\text{if $L=2$;}
  \\
  \e^{\mp\i\pi(\ve_2+\ve_3+1)},
  &\text{if $L=3$,}
}
\label{BMS-fin}
\\
F^{MS}_{\alpha\beta}\sbm{2&3\\1&4}
&=\Cases{{\sin\pi(\ve_1+\ve_2)\over\sin\pi(\ve_3+\ve_2)},
  &\text{if $L=1$;}
  \\
  F_{\alpha\beta}(\ve_1,\ve_2,\ve_3),
  &\text{if $L=2$;}
  \\
  1,
  &\text{if $L=3$.}
}
\label{FMS-fin}
$$
Equations (\ref{hex})--(\ref{BeqF}) for these matrices were also checked by a straightforward calculation.

%%%%%%%%%%%%%%%%%%%%%%%%%%%%%%%%%%%%%%%%%%%%%%%%%%%%%%%%%%%%%%%%%%%%%%%%
\section{Special modules and logarithmic operators}
\label{sec-logops}

%%%%%%%%%%%%%%%%%%%%%%%%%%%%%%%%%%%%%%%%%%%%%%%%%%%%%%%%%%%%%%%%%%%%%%%%

\subsection{Special modules}
\label{subsec-specmod}

Here we discuss special points of the parameter $e$, where the Verma modules $M^l_{en}$ degenerate. It is easy to see that if
\eq$$
\Delta^l_{e,n-1/2}=\Delta^{l+1}_{e,n+1/2},
\label{DeltaDelta-eq}
$$
which is equivalent to
\eq$$
e=-l\in\half\Z,
\label{le-rel}
$$
the Verma modules $M^l_{e,n-1/2}$ and $M^{l+1}_{e,n+1/2}$ are isomorphic and can be identified.%
\footnote{The corresponding Fock modules $\cF^l_{e,n-1/2}$ and $\cF^{l+1}_{e,n+1/2}$ have conjugate structures.}
We denote this module~$M^\circ_{en}$ and the corresponding irreducible module $\cV^\circ_{en}$. It contains a highest weight vector
\eq$$
|v^\circ_{en}\rangle=|v^{-e}_{e,n-1/2}\rangle=|v^{1-e}_{e,n+1/2}\rangle,
\label{vcirc-def}
$$
which satisfies
\eq$$
L_0|v^\circ_{en}\rangle=\Delta^\circ_{en}|v^\circ_{en}\rangle,
\qquad
J^E_0|v^\circ_{en}\rangle=e|v^\circ_{en}\rangle,
\qquad
J^N_0|v^\circ_{en}\rangle=(e+n)|v^\circ_{en}\rangle
\label{vcirc-zeromodes}
$$
with the Virasoro weight
\eq$$
\Delta^\circ_{en}=\Delta^{-e}_{e,n-1/2}=\Delta^{1-e}_{e,n+1/2}=e(e+n).
\label{DeltaO-en-def}
$$
This module is degenerate and contains null vectors at the levels ${1\over2}k(k+1)$ ($k=1,2,\ldots$).

\begin{figure}[t]
\center
\begin{tikzpicture}[scale=1.5,font=\small,bullet/.style={inner sep=2pt,outer sep=1pt,circle,fill}]
\foreach \x in {0,1} \node[bullet] (V \x\space 0) at (\x,0) {};
\foreach \x in {-1,...,2} \node[bullet] (V \x\space 1) at (\x,-1) {};
\foreach \x in {-2,...,3} \node[bullet] (V \x\space 2) at (\x,-2) {};
\foreach \x in {-3,...,4} \node[bullet] (V \x\space 3) at (\x,-3) {};
\node[below] at (-3,-3.5) {$\cV^{-e-3}_{e,n-{7\over2}}$};
\node[below] at (-2,-3.5) {$\cV^{-e-2}_{e,n-{5\over2}}$};
\node[below] at (-1,-3.5) {$\cV^{-e-1}_{e,n-{3\over2}}$};
\node[below] at (0,-3.5) {$\cV^{-e}_{e,n-{1\over2}}$};
\node[below] at (1,-3.5) {$\cV^{-e+1}_{e,n+{1\over2}}$};
\node[below] at (2,-3.5) {$\cV^{-e+2}_{e,n+{3\over2}}$};
\node[below] at (3,-3.5) {$\cV^{-e+3}_{e,n+{5\over2}}$};
\node[below] at (4,-3.5) {$\cV^{-e+4}_{e,n+{7\over2}}$};
\node[above,xshift=-6pt] at (V -3 3) {$u^{-3}_0$};
\node[above,xshift=-6pt] at (V -2 2) {$u^{-2}_0$}; \node[right,yshift=4pt] at (V -2 3) {$w^{-2}_0$};
\node[above,xshift=-6pt] at (V -1 1) {$u^{-1}_0$}; \node[right,yshift=4pt] at (V -1 2) {$w^{-1}_0$};
  \node[left,yshift=-4pt] at (V -1 3) {$u^{-1}_1$};
\node[above,xshift=-6pt] at (V 0 0) {$u^0_0$}; \node[right,yshift=4pt] at (V 0 1) {$w^0_0$}; \node[left,yshift=-4pt] at (V 0 2) {$u^0_1$};
  \node[right,yshift=4pt] at (V 0 3) {$w^0_1$};
\node[above,xshift=8pt] at (V 1 0) {$w^1_0$}; \node[left,yshift=-4pt] at (V 1 1) {$u^1_0$}; \node[right,yshift=4pt] at (V 1 2) {$w^1_1$};
  \node[left,yshift=-4pt] at (V 1 3) {$u^1_1$};
\node[above,xshift=8pt] at (V 2 1) {$w^2_0$}; \node[left,yshift=-4pt] at (V 2 2) {$u^2_0$}; \node[right,yshift=4pt] at (V 2 3) {$w^2_1$};
\node[above,xshift=8pt] at (V 3 2) {$w^3_0$}; \node[left,yshift=-4pt] at (V 3 3) {$u^3_0$};
\node[above,xshift=8pt] at (V 4 3) {$w^4_0$};
\foreach \y in {0,2} \draw[<-] (V 0 \y) -- node[above] {$\cQ$} (V 1 \y);
\foreach \y in {1,3} {\draw[<-] (V -1 \y) -- node[above] {$\cQ$} (V 0 \y); \draw[<-] (V 1 \y) -- node[above] {$\cQ$} (V 2 \y);}
\draw[<-] (V -2 2) -- node[above] {$\cQ$} (V -1 2); \draw[<-] (V 2 2) -- node[above] {$\cQ$} (V 3 2);
\draw[<-] (V -3 3) -- node[above] {$\cQ$} (V -2 3); \draw[<-] (V 3 3) -- node[above] {$\cQ$} (V 4 3);
\draw[<-] (V -2 2) -- (V -2 3);
\draw[<-] (V -1 1) -- (V -1 2); \draw[->] (V -1 2) -- (V -1 3);
\draw[<-] (V 0 0) -- (V 0 1); \draw[->] (V 0 1) -- (V 0 2); \draw[<-] (V 0 2) -- (V 0 3);
\draw[->] (V 1 0) -- (V 1 1); \draw[<-] (V 1 1) -- (V 1 2); \draw[->] (V 1 2) -- (V 1 3);
\draw[->] (V 2 1) -- (V 2 2); \draw[<-] (V 2 2) -- (V 2 3);
\draw[->] (V 3 2) -- (V 3 3);
\foreach \x in {-3,-1,1,3} \draw[<-] (V \x\space 3) -- (\x,-3.5);
\foreach \x in {-2,0,2,4} \draw[-] (V \x\space 3) -- (\x,-3.5);
\end{tikzpicture}
\caption{Action of the screening operator $\cQ$ on the degenerate modules of the algebra $\Vir^{EN}$ for $e\in{1\over2}\Z$. The letter $u^\ve_s$ ($\ve\in\Z$) denotes the vector $|u^\ve_{e,n+\ve;s}\rangle\in\cV^{-e+\ve}_{e,n-1/2+\ve}$, which is either the highest weight vector or a singular vector, while the letter $w^\ve_s$ denotes the vector $|w^\ve_{e,n+\ve;s}\rangle\in\cV^{-e+\ve}_{e,n-1/2+\ve}$, which is either the highest weight vector or a cosingular vector (defined modulo $\Ker\cQ$).}
\label{fig-felder}
\end{figure}

The structure of the corresponding Fock module is defined by action of the screening operator, according to the Felder\-/type diagram on Fig.~\ref{fig-felder}. This diagram shows that if $\ve=e+l\in\Z$, there is a resonance between the vector $|u^\ve_0\rangle$ and $|w^{\ve+1}_0\rangle$ in the notation of the Figure. One of these vectors is a highest weight one, while the other is a cosingular or singular vector except the case $\ve=0$. These vectors for nonzero $\ve$ can be obtained by recursive relations
\eq$$
\Aligned{
|u^\ve_{en;0}\rangle
&=J^+_{\ve-e}|u^{\ve+1}_{en;0}\rangle,
\quad
&|w^{\ve+1}_{en;0}\rangle
&=J^+_{\ve-e}|w^{\ve+2}_{en;0}\rangle
&&\quad\text{for $\ve<0$;}
\\
|u^\ve_{en;0}\rangle
&=\ve^{-1}J^-_{e-\ve}|u^{\ve-1}_{en;0}\rangle,
\quad
&|w^{\ve+1}_{en;0}\rangle
&=\ve^{-1}J^-_{e-\ve}|w^\ve_{en;0}\rangle
&&\quad\text{for $\ve>0$.}
\\
}\label{vw-recursions}
$$
The equation for $w^\ve_0$ in the first line and for $v^\ve_0$ in the second line are obtained from the commutativity $[Q,J^\pm_r]=0$. Note that the action of the operators $J^\pm_r$ is transversal to the diagram on Fig.~\ref{fig-felder}: it does not change the $n$ parameter, while the $Q$ operator changes it. Below we concentrate on the case $\ve=0$, while the case of integer $\ve\ne0$ will be discussed at the end of the section.

Now return to the resonance (\ref{DeltaDelta-eq}). The formal limit
\eq$$
|v'_{en}\rangle=\lim_{\delta\to0}{|v^{-e}_{e+\delta,n-1/2}\rangle-|v^{1-e}_{e+\delta,n+1/2}\rangle\over\delta}
\label{vprime-def}
$$
satisfies the relations
\eq$$
\Gathered{
L_k|v'_{en}\rangle=0,
\qquad
J^{E,N}_k|v'_{en}\rangle=0
\quad(k>0),
\\
L_0|v'_{en}\rangle=\Delta^\circ_{en}|v'_{en}\rangle-|v^\circ_{en}\rangle,
\qquad
J^E_0|v'_{en}\rangle=e|v'_{en}\rangle,
\qquad
J^N_0|v'_{en}\rangle=(e+n)|v'_{en}\rangle.
}\label{vprime-Laction}
$$
The operators $L_0$ acts on the pair $(|v^\circ_{en}\rangle,-|v'_{en}\rangle)$ as a Jordanian cell. By acting on this vector the algebra $\Vir^{EN}$ generates a new module $M'_{en}\supset M^\circ_{en}$. The structure of this module is not the subject of this paper, but we will briefly discuss the scalar product in it.

The conjugate vectors should be defined as formal limits
\eq$$
\langle v^\circ_{en}|=\lim_{\delta\to0}\delta\langle v^{-e}_{e+\delta,n-1/2}|=-\lim_{\delta\to0}\delta\langle v^{1-e}_{e+\delta,n+1/2}|,
\qquad
\langle v'_{en}|=\lim_{\delta\to0}\left(\langle v^{-e}_{e+\delta,n-1/2}|+\langle v^{1-e}_{e+\delta,n+1/2}|\right).
\label{vcirc-vprime-conj}
$$
With this definition we have the scalar products
\eq$$
\langle v^\circ_{e'n'}|v^\circ_{en}\rangle=\langle v'_{e'n'}|v'_{en}\rangle=0,
\qquad
\langle v^\circ_{e'n'}|v'_{en}\rangle=\langle v'_{e'n'}|v^\circ_{en}\rangle=\delta_{e'e}\delta_{n'n},
\label{vcirc-vprime-scal}
$$
if we assume~(\ref{vlen-norm}). We set the same value of $\delta$ in the definitions (\ref{vcirc-def}), (\ref{vprime-def}), (\ref{vcirc-vprime-conj}) and took the limit $\delta\to0$ in the scalar products.

Then let us find the Shapovalov form on the module~$M'_{en}$. Let us start from the Shapovalov forms on the modules $\cV^l_{en}$ for generic values of~$e$. Let $|K|=|K'|$. Then
\eq$$
W_{K'K}(e,n+{\textstyle{1\over2}}-l,\Delta^l_{en})=\langle v^l_{en}|\Lambda_{K'}\Lambda_{-K}|v^l_{en}\rangle.
\label{WK'K-def}
$$
Since $\Lambda_K$ is a product of the operators $J^E_k$, $J^N_k$, $L_k$, the Shapovalov form depends on the eigenvalues of the zero modes $J^E_0$, $J^N_0$, $L_0$, as it is indicated here.

Since the matrix elements $W_{KK'}$ are vanish for $|K|\ne|\bar K|$, this matrix is block\-/diagonal with finite\-/dimensional blocks. We shall denote the block for $N=|K|=|K'|$ by~$W_N$. Similar notation will be used below for derived matrices: $W^\circ_N$ etc.

Now assume (\ref{le-rel}) and calculate matrix elements in the module $M'_{en}$ similarly to (\ref{vcirc-vprime-scal}). Evidently, the $\langle v^\circ|v^\circ\rangle$ type scalar products vanish:
\eq$$
\langle v^\circ_{en}|\Lambda_{K'}\Lambda_{-K}|v^\circ_{en}\rangle=\langle v^\circ_{en}|\Lambda_{K'}\Lambda_{-K}(\Delta^\circ_{en}-L_0)|v'_{en}\rangle
=\langle v^\circ_{en}|(\Delta^\circ_{en}-L_0)\Lambda_{K'}\Lambda_{-K}|v'_{en}\rangle=0,
\label{vcirc-vcirc-zero}
$$
if $|K|=|K'|$. For $\langle v^\circ|v'\rangle$ type scalar products we have
\eq$$
\langle v'_{en}|\Lambda_{K'}\Lambda_{-K}|v^\circ_{en}\rangle=\langle v^\circ_{en}|\Lambda_{K'}\Lambda_{-K}|v'_{en}\rangle
=W_{K'K}(e,n+e,\Delta^\circ_{en})\equiv W^\circ_{K'K}(e,n).
\label{vprime-vcirc-WK'K}
$$

At last, for the $\langle v'|v'\rangle$ type scalar products we obtain
\Multline$$
\langle v'_{en}|\Lambda_{K'}\Lambda_{-K}|v'_{en}\rangle
=\lim_{\delta\to0}\delta^{-1}\left(\langle v^{-e}_{e+\delta,n-1/2}|+\langle v^{1-e}_{e+\delta,n+1/2}|\right)\Lambda_{K'}\Lambda_{-K}
\left(|v^{-e}_{e+\delta,n-1/2}\rangle-|v^{1-e}_{e+\delta,n+1/2}\rangle\right)
\\
=\lim_{\delta\to0}\delta^{-1}\left(W_{K'K}(e+\delta,n+e,\Delta^{-e}_{e+\delta,n-1/2})-W_{K'K}(e+\delta,n+e,\Delta^{1-e}_{e+\delta,n+1/2})\right)
\\
=-\left.{\d\over\d\Delta}W_{K'K}(e,n+e,\Delta)\right|_{\Delta=\Delta^\circ_{en}}
\equiv W'_{K'K}(e,n).
\label{vprime-vprime-zero}
$$

Thus we may define a total Shapovalov form. In the basis of the form $(\Lambda_{-K}|v^\circ_{en}\rangle,\Lambda_{-K}|v'_{en}\rangle)$ it looks like
\eq$$
\hat W=\pMatrix{0&W^\circ\\W^\circ&W'}.
\label{hatW-def}
$$
For $N\ge1$ the blocks $W^\circ_N$ are degenerate. Hence, the blocks $\hat W_N$ are degenerate as well.

To understand the situation consider the simplest case $N=1$. We have
$$
W^\circ_1(e,n)=\pMatrix{2e(e+n)&e&e+n\\e&0&1\\e+n&1&0},
\qquad
W'_1(e,n)=\pMatrix{-2&0&0\\0&0&0\\0&0&0}.
$$
in the basis $(L_{-1},J^E_{-1},J^N_{-1})$. The matrix $W^\circ_1$ has corank~$1$, which corresponds to the only null vector on this level:
$$
|\chi_{en}\rangle=(L_{-1}-(e+n)J^E_{-1}-eJ^N_{-1})|v^\circ_{en}\rangle.
$$
This is the only first level null vector in the whole module $M'_{en}$, and other null vectors belong to the submodule $M^1_{en}=\Vir^{EN}|\chi_{en}\rangle$. Hence, the irreducible module is
\eq$$
\cV'_{en}=M'_{en}/M^1_{en}.
\label{cV'-def}
$$
There is a natural map from $M^\circ_{en}$ to $M'_{en}\supset M^\circ_{en}$, which maps the `usual' states onto the `logarithmic' ones:
\eq$$
\cI:\Lambda_{-K}|v^\circ_{en}\rangle\mapsto\Lambda_{-K}|v'_{en}\rangle.
\label{cImap-def}
$$
This map provides an isomorphism $M^\circ_{en}\simeq M'_{en}/M^\circ_{en}$. It will be useful later.

Return to the level 1 subspace. The matrix $\hat W_1$ is a corank~$1$ matrix as well. In what follows we will need to invert such kind of matrices to obtain decompositions of unity. We could use the Moore\--Penrose pseudoinverse, but there is an easier way. Let us drop out the any basic element in the $\cV^\circ_{en}$ submodule, e.g. $L_{-1}|v^\circ_{en}\rangle$. It corresponds to dropping one raw in~$W^\circ$:
$$
\tilde W^\circ_1(e,n)=\pMatrix{e&0&1\\e+n&1&0}.
$$
In the basis $(J^E_{-1}|v^\circ\rangle,J^N_{-1}|v^\circ\rangle,L_{-1}|v'\rangle,J^E_{-1}|v'\rangle,J^N_{-1}|v'\rangle)$ we construct
$$
\tilde W_1=\pMatrix{0&\tilde W^\circ_1\\\tilde W^{\circ T}_1&W'_1}
=\left(\begin{array}{cc|ccc}0&0&e&0&1\\0&0&e+n&1&0\\\hline e&e+n&-2&0&0\\0&1&0&0&0\\1&0&0&0&0\end{array}\right),
$$
which is nondegenerate and can be inverted. More generally, on any level $N$ drop out the number of basic elements equal to the corank of $W^\circ_N$ and define $\tilde W^\circ_N$, which differs by omitting the corresponding rows. Then the matrix
$$
\tilde W_N=\pMatrix{0&\tilde W^\circ_N\\\tilde W^{\circ T}_N&W'_N}
$$
is nondegenerate and can be inverted.

Now turn to the corresponding physical operators in the conformal field theory.

%%%%%%%%%%%%%%%%%%%%%%%%%%%%%%%%%%%%%%%%%%%%%%%%%%%%%%%%%%%%%%%%%%%%%%%%
\subsection{Primary logarithmic operators}

Recall the definition of the primary logarithmic operators at special points~\cite{Schomerus:2005bf}. Consider the structure constants corresponding to two regular operators and one in the vicinity of the degeneracy point. Namely, let $e_1=e'_1=e+\delta$ ($e\in\half\Z$), $l_1=\bl_1=l'_1-1=\bl'_1-1=1-l_2-l_3=1-\bl_2-\bl_3=-e$ and consider the limit $\delta\to0$. It is straightforward to find from (\ref{C123-final}) that
\eq$$
{C_{1'23}\over C_{123}}=-\e^{\i\pi e}(\delta^{-1}+D_3)+O(\delta).
\label{C123-degen-ratio}
$$
The first correction $D_3=D^{l_3\bl_3}_{e_3}$ will be needed below. It reads
\eq$$
D^{l\bl}_e=\psi(e+l)+\psi(1-e-\bl)-2\psi(1).
\label{D1-def}
$$
Here $\psi(u)=(\log\Gamma(u))'$ is the digamma function. It is easy to check that $D_{i^*}=D_i$ and
\eq$$
D_2=D_3,
\quad\text{if $e_2+e_3+e=0$, $l_2+l_3=\bl_2+\bl_3=1+e$.}
\label{D2D1-id}
$$
The leading contribution to the ratio (\ref{C123-degen-ratio}) is independent of the operators~$\Phi_2$, $\Phi_3$. This gives us a hint that the operators $\Phi^{-e,-e}_{e+\delta,n-1/2}(x)$ and $(-\e^{\i\pi e}\delta)\Phi^{1-e,1-e}_{e+\delta,n+1/2}(x)$ should be identified in the limit $\delta\to0$. To complete this identification we have to compare conformal blocks corresponding these operators. We postpone the discussion of the conformal blocks to section~\ref{sec-cb-degen}. Now let us fix the normalization. To do it let us calculate the $\delta\to0$ limit of the structure constants:
\Align$$
C^\circ_{123}
\equiv\lim_{\delta\to0}C_{123}\gamma_e(\delta)
=\Gamma_{123}\left(\Gamma(1-\{l_2\}+e_2)\Gamma(1-\{l_3\}+e_3)
  \over\Gamma(1-\{l_2\}-e_2)\Gamma(1-\{l_3\}-e_3)\right)^{1/2},
\label{limC123}
$$
where
\eq$$
\gamma_e(\delta)=\left(\Gamma(1-\{e\}-e-\delta)\over\Gamma(1-\{e\}+e+\delta)\right)^{1/2}.
\label{gamma-e-def}
$$
and
\eq$$
\Gamma_{123}=\e^{\i\pi\left(l_2\bl_3-e(e+1)\right)}
\label{Gamma123-special}
$$
It is easy to check that $\gamma_e(\delta)\sim\delta^{-{1\over2}\sign e}(1+O(\delta))$ for $e\in\half\Z$. Thus we may define an operator
\eq$$
\Phi^\circ_{en}(x)
=\lim_{\delta\to0}\gamma_e(\delta)\Phi^{-e,-e}_{e+\delta,n-1/2}(x)
=-\e^{-\i\pi e}\lim_{\delta\to0}\delta\gamma_e(\delta)\Phi^{1-e,1-e}_{e+\delta,n+1/2}(x),
\label{PhiO-def}
$$
Define also an operator, which is the difference of these two for finite~$\delta$:
\eq$$
\Psi_{en,\delta}(x)
=\gamma_e(\delta)\left(\delta^{-1}\Phi^{-e,-e}_{e+\delta,n-1/2}(x)
  +\e^{-\i\pi e}\Phi^{1-e,1-e}_{e+\delta,n+1/2}(x)\right).
\label{Psi-delta-def}
$$
Its limit
\eq$$
\Psi_{en}(x)=\lim_{\delta\to0}\Psi_{en,\delta}(x)
\label{Psi-def}
$$
provides a logarithmic operator in the $\widehat{gl}(1|1)$ model. Indeed, we easily prove
\eq$$
\Aligned{
L_0|\Phi^\circ_{en}\rangle
&=\Delta^\circ_{en}|\Phi^\circ_{en}\rangle,
&L_r|\Phi^\circ_{en}\rangle
&=J^{E,N}_r|\Phi^\circ_{en}\rangle=0,
\\
L_0|\Psi_{en}\rangle
&=\Delta^\circ_{en}|\Psi_{en}\rangle-|\Phi^\circ_{en}\rangle,
&L_r|\Psi_{en}\rangle
&=J^{E,N}_r|\Psi_{en}\rangle=0\quad(r>0),
\\
J^E_0|\Phi^\circ_{en}\rangle
&=e|\Phi^\circ_{en}\rangle,
&J^N_0|\Phi^\circ_{en}\rangle
&=(n+e)|\Phi^\circ_{en}\rangle,
\\
J^E_0|\Psi_{en}\rangle
&=e|\Psi_{en}\rangle,
&J^N_0|\Psi_{en}\rangle
&=(n+e)|\Psi_{en}\rangle.
}\label{L0-PhiO-Psi-action}
$$
The action of $\bar L_r$ ($r\ge0$) is the same. We see that both the operators $L_0$ and $\bar L_0$ act on the pair $(|\Phi^\circ_{en}\rangle,-|\Psi_{en}\rangle)$ by a Jordanian cell. It means that we should identify
\eq$$
|\Phi^\circ_{en}\rangle\sim|v^\circ_{en}\rangle\otimes|v^\circ_{en}\rangle,
\qquad
|\Psi_{en}\rangle\sim|v'_{en}\rangle\otimes|v^\circ_{en}\rangle+|v^\circ_{en}\rangle\otimes|v'_{en}\rangle.
\label{PhicircPsi-vcircv'-rel}
$$
The subspace of states over $|\Psi_{en}\rangle$ as a module of $\Vir^{EN}\otimes\Vir^{EN}$ is the submodule
$$
\hat\cV_{en}=\Vir^{EN}\otimes\Vir^{EN}(|v'_{en}\rangle\otimes|v^\circ_{en}\rangle+|v^\circ_{en}\rangle\otimes|v'_{en}\rangle)
$$
in the product $\cV'_{en}\otimes\cV'_{en}$.

It is easy to see that the norm of the vector $|\Phi^\circ_{en}\rangle$ vanishes:
\eq$$
\langle\Phi^\circ_{e'n'}|\Phi^\circ_{en}\rangle=0.
\label{PhiO-PhiO-scalprod}
$$
Indeed,
$$
\Gathered{
(\Delta^\circ_{en}-\Delta^\circ_{e'n'})\langle\Phi^\circ_{e'n'}|\Phi^\circ_{en}\rangle
=\langle\Phi^\circ_{e'n'}|(L_0-L_0)|\Phi^\circ_{en}\rangle=0,
\\
\langle\Phi^\circ_{-e,-n}|\Phi^\circ_{en}\rangle
=\langle\Phi^\circ_{-e,-n}|(\Delta^\circ_{en}-L_0)|\Psi_{en}\rangle=0.
}
$$
But, in contrast to the null vectors, these states are not orthogonal to all other states. As we will see below $\langle\Phi_{-e,-n}|\Psi_{en}\rangle\ne0$.

Now let us discuss the unit operator and the vacuum state in the model, which correspond to $e=n=0$. We have $C^\circ_{123}=\e^{\i\pi l_2(1-\bl_2)}$, which is consistent with the normalization condition~(\ref{PhiPhi-corr}), if we assume
\eq$$
\Phi^\circ_{00}(x)=1.
\label{Phi-unit}
$$
Due to (\ref{PhiO-PhiO-scalprod}) the vacuum state in the theory has a zero norm:
\eq$$
\langle\vac|\vac\rangle=0.
\label{vac-zeronorm}
$$

Consistently with (\ref{PhiO-PhiO-scalprod}) any $\langle\Phi^\circ\Phi^\circ\rangle$ correlation function vanishes:
\Multline$$
\langle\Phi^\circ_{e'n'}(x')\Phi^\circ_{en}(x)\rangle
=-\e^{-\i\pi e}\lim_{\delta\to0}\delta\gamma_{e'}(-\delta)\gamma_e(\delta)
\langle\Phi^{-e',-e'}_{e'-\delta,n'-1/2}(x')\Phi^{1-e,1-e}_{e+\delta,n+1/2}(x)\rangle
\\
=-\e^{-\i\pi e^2}\delta_{e+e',0}\delta_{n+n',0}\lim_{\delta\to0}\delta|z'-z|^{-4\Delta^{1-e}_{e+\delta,n+1/2}}=0.
\label{PhiO-PhiO-corr}
$$
Consider three\-/point functions. Let $l'_2=l_2+1$, $\bl'_2=\bl_2+1$, $n'_2=n_2+1$. Then we obtain
\eq$$
\lim_{\delta\to0}C_{12'3}\gamma_e(\delta)=0.
\label{C12'3-zero}
$$
This means that a correlation function of the form $\langle\Phi\Phi\Phi^\circ\rangle$ vanishes for $\sum n_i=1$. Thus we have
\Align$$
\langle\Phi_3(x_3)\Phi_2(x_2)\Phi^\circ_{en}(x_1)\rangle
&=C^\circ_{123}\delta_{e+e_2+e_3,0}\delta_{n+n_2+n_3,0}\delta_{l_2+l_3,1+e}\delta_{\bl_2+\bl_3,1+e}
\notag
\\*
&\quad\times\prod^3_{i<j\atop k\ne i,j}(z_j-z_i)^{\Delta_k-\Delta_i-\Delta_j}(\bz_j-\bz_i)^{\bDelta_k-\bDelta_i-\bDelta_j},
\label{Phi12PhiO3-corr}
$$
where $\Delta_1=\bDelta_1=\Delta^\circ_{en}$.

Now consider the correlation functions that contain the $\Psi$ operators. First of all, from
\Multline*$$
\gamma_{e'}(-\delta)\langle\Phi^{-e',-e'}_{e'-\delta,n'-1/2}(x')\Psi_{en,\delta}(x)\rangle
=\e^{-\i\pi e}\langle\Phi^{-e',-e'}_{e'-\delta,n'-1/2}(x')\Phi^{1-e,1-e}_{e+\delta,n+1/2}(x)\rangle
\\*
=\e^{-\i\pi e^2}\delta_{e+e',0}\delta_{n+n',0}|z'-z|^{-4\Delta^{1-e}_{e+\delta,n+1/2}},
$$
we immediately obtain
\eq$$
\langle\Phi^\circ_{e'n'}(x')\Psi_{en}(x)\rangle=\e^{-\i\pi e^2}\delta_{e+e',0}\delta_{n+n',0}|z'-z|^{-4\Delta^\circ_{en}}.
\label{PhiO-Psi-corr}
$$
All other $\langle\Phi\Psi\rangle$ correlation functions vanish. If we define
\eq$$
\langle\Phi^\circ_{en}|=\langle\vac|e^{\i\pi e^2}\Phi^\circ_{-e,-n}(\infty,\infty),
\qquad
\Phi^\circ_{en}(\infty,\infty)=\lim_{\zeta\to\infty}|\zeta|^{4\Delta^\circ_{en}}\Phi^\circ_{en}(\zeta,\bar\zeta),
\label{Phicirc-bra-def}
$$
we obtain
\eq$$
\langle\Phi^\circ_{e'n'}|\Psi_{en}\rangle=\delta_{e'e}\delta_{n'n}.
\label{PhiO-Psi-scalprod}
$$
In particular, we have the vacuum expectation value
\eq$$
\langle\Psi_{en}(x)\rangle=\delta_{e0}\delta_{n0}.
\label{Psi-VEV}
$$
In a similar way as (\ref{PhiO-Psi-corr}) we derive
\eq$$
\langle\Psi_{e'n'}(x')\Psi_{en}(x)\rangle=2\langle\Phi^\circ_{e'n'}(x')\Psi_{en}(x)\rangle\log|z'-z|^2.
\label{Psi-Psi-corr}
$$
For the tree\-/point functions we have
\eq$$
\langle\Phi_3(x_3)\Phi_2(x_2)\Psi_{en}(x_1)\rangle
=\langle\Phi_3(x_3)\Phi_2(x_2)\Phi^\circ_{en}(x_3)\rangle
\left(-D_3+\log{|z_2-z_1|^2|z_3-z_1|^2\over|z_3-z_2|^2}\right).
\label{Psi-Phi-Phi-corr}
$$

Note that the definition (\ref{Psi-def}) is not unique. The relations (\ref{L0-PhiO-Psi-action}) are invariant under the transformation
\eq$$
\Psi_{en}(x)\to\Psi_{en}(x)+\const\cdot\Phi^\circ_{en}(x).
\label{Psi-trans}
$$
Nevertheless, such transformation is equivalent to adding extra constant factors under the logarithms. Surely, it is impossible to choose a definition, which could completely cancel the $D_3$ contribution in all $\langle\Psi\Phi\Phi\rangle$ correlation functions.

The formulas (\ref{Psi-Psi-corr}) and (\ref{Psi-Phi-Phi-corr}) with the invariance (\ref{Psi-trans}) give us a hint how to define the state $\langle\Psi_{en}|$. Namely, define
\eq$$
\Psi_{en}(\infty,\infty)=\lim_{\zeta\to\infty}|z|^{4\Delta^\circ_{en}}\left(\Psi_{en}(\zeta,\bar\zeta)-2\Phi^\circ_{en}(\zeta,\bar\zeta)\log|\zeta|^2\right).
\label{Psi-inf-def}
$$
Then
\eq$$
\langle\Psi_{en}|=\langle\vac|\e^{i\pi e^2}\Psi_{-e,-n}(\infty,\infty).
\label{Psi-bra-def}
$$
With these definitions we have
\eq$$
\langle\Psi_{e'n'}|\Psi_{en}\rangle=0,
\qquad
\langle\Psi_{e'n'}|\Phi^\circ_{en}\rangle=\delta_{e'e}\delta_{n'n}
\label{Psi-Psi-matel}
$$
and
\eq$$
\Aligned{
\langle\Psi_{en}|\Phi_2(z,\bz)|\Phi_3\rangle
&=\langle\Phi^\circ_{en}|\Phi_2(z,\bz)|\Phi_3\rangle\left(-D_3-\log|z|^2\right),
\\
\langle\Phi_3|\Phi_2(z,\bz)|\Psi_{en}\rangle
&=\langle\Phi_3|\Phi_2(z,\bz)|\Phi^\circ_{en}\rangle\left(-D_3+\log|z|^2\right).
}
\label{Psi-Phi-Phi-matel}
$$
We will use these equations later when discussing the $s$\-/channel decomposition of four\-/point correlation functions.

%%%%%%%%%%%%%%%%%%%%%%%%%%%%%%%%%%%%%%%%%%%%%%%%%%%%%%%%%%%%%%%%%%%%%%%%
\subsection{Descendant logarithmic operators}

Now consider the action of $\Vir^{EN}$ on the vectors $|\Phi^{-e,-e}_{e+\delta,n-1/2}\rangle$ and $|\Phi^{1-e,1-e}_{e+\delta,n+1/2}\rangle$ in the limit $\delta\to0$.

First of all, notice that every matrix element $\langle\Phi_3|\Phi_2(z)\Lambda_{-K}\bar\Lambda_{-\bar K}|\Phi_1\rangle$ factorizes as follows:
\eq$$
\langle\Phi_3|\Phi_2(z)\Lambda_{-K}\bar\Lambda_{-\bar K}|\Phi_1\rangle
=z^{\Delta_3-\Delta_2-\Delta_1-|K|}\bz^{\bar\Delta_3-\bar\Delta_2-\bar\Delta_1-|\bar K|}f_{2,K}\bar f_{2,\bar K}C_{12}{}^3,
\label{3point-factorization}
$$
where the factor $f_{2,K}$ only depends on $K$, $l_2$, $e_2$, $n_2$ and the factor $\bar f_{2,\bar K}$ is the same with the substitution $l_2\to\bl_2$, $K\to\bar K$. Indeed, the factor $f_{2,K}$ originates in the commutators of $\Phi_2(z)$ with the currents $J^E_{-k}$, $J^N_{-k}$ and~$L_{-k}$. It means that the following limits are well\-/defined:
\eq$$
\lim_{\delta\to0}\gamma_e(\delta)\Lambda_{-K}\bar\Lambda_{-\bar K}|\Phi^{-e,-e}_{e+\delta,n-1/2}\rangle
=-\e^{-\i\pi e}\lim_{\delta\to0}\delta\gamma_e(\delta)\Lambda_{-K}\bar\Lambda_{-\bar K}|\Phi^{1-e,1-e}_{e+\delta,n+1/2}\rangle
=\Lambda_{-K}\bar\Lambda_{-\bar K}|\Phi^\circ_{en}\rangle
\label{PhiO-state-descendant}
$$
and
\eq$$
\lim_{\delta\to0}\Lambda_{-K}\bar\Lambda_{-\bar K}|\Psi_{en,\delta}\rangle=\Lambda_{-K}\bar\Lambda_{-\bar K}|\Psi_{en}\rangle.
\label{Psi-state-descendant}
$$
The descendant operators $(\Lambda_{-K}\bar\Lambda_{-\bar K}\Phi^\circ_{en})(z,\bz)$ and $(\Lambda_{-K}\bar\Lambda_{-\bar K}\Psi_{en})(z,\bz)$ are defined by the standard operator\--state correspondence.

It is easy to see that the operator $L_0$ has a Jordan block on the states defined in (\ref{PhiO-state-descendant}), (\ref{Psi-state-descendant}):
\eq$$
\Aligned{
L_0\Lambda_{-K}\bar\Lambda_{-\bar K}|\Phi^\circ_{en}\rangle
&=(\Delta^\circ_{en}+|K|)\Lambda_{-K}\bar\Lambda_{-\bar K}|\Phi^\circ_{en}\rangle,
\\
L_0\Lambda_{-K}\bar\Lambda_{-\bar K}|\Psi_{en}\rangle
&=(\Delta^\circ_{en}+|K|)\Lambda_{-K}\bar\Lambda_{-\bar K}|\Psi_{en}\rangle-\Lambda_{-K}\bar\Lambda_{-\bar K}|\Phi^\circ_{en}\rangle.
}\label{L0-PhiO-Psi-descendant-action}
$$

By generalizing calculations of the last subsection, we obtain the matrix elements
\Align$$
\langle\Phi^\circ_{e'n'}|\Lambda_{K'}\bar\Lambda_{\bar K'}\Lambda_{-K}\bar\Lambda_{-K}|\Phi^\circ_{en}\rangle
&=0,
\label{Phicirc-Lambda-Phicirc-matel}
\\
\langle\Psi_{e'n'}|\Lambda_{K'}\bar\Lambda_{\bar K'}\Lambda_{-K}\bar\Lambda_{-K}|\Phi^\circ_{en}\rangle
&=\langle\Phi^\circ_{e'n'}|\Lambda_{K'}\bar\Lambda_{\bar K'}\Lambda_{-K}\bar\Lambda_{-K}|\Psi_{en}\rangle
=\delta_{e'e}\delta_{n'n}W_{K'K}(e,n)W_{\bar K'\bar K}(e,n),
\label{Psi-Lambda-Phicirc-matel}
\\
\langle\Psi_{e'n'}|\Lambda_{K'}\bar\Lambda_{\bar K'}\Lambda_{-K}\bar\Lambda_{-K}|\Psi_{en}\rangle
&=\delta_{e'e}\delta_{n'n}\left(W^\circ_{K'K}(e,n)W'_{\bar K'\bar K}(e,n)+W'_{K'K}(e,n)W^\circ_{\bar K'\bar K}(e,n)\right).
\label{Psi-Lambda-Psi-matel}
$$
In other words, the metric on the space of states in the basis is given by the matrix
\eq$$
\hat U=\pMatrix{0&W^\circ\otimes W^\circ\\W^\circ\otimes W^\circ&W^\circ\otimes W'+W'\otimes W^\circ}
\label{hatU-def}
$$
in the $(|\Phi^\circ\rangle,|\Psi\rangle)$ basis. Due to degeneracy of the matrix $W^\circ_N$ for $N\ge1$ this matrix is also degenerate. We can define a nondegenerate matrix
\eq$$
\tilde U_N=\pMatrix{0&\tilde W^\circ\otimes\tilde W^\circ\\\tilde W^{\circ T}\otimes\tilde W^{\circ T}&W^\circ\otimes W'+W'\otimes W^\circ}.
\label{tildeU-def}
$$
This defines the all\-/level form $\tilde U(e,n)$.

%%%%%%%%%%%%%%%%%%%%%%%%%%%%%%%%%%%%%%%%%%%%%%%%%%%%%%%%%%%%%%%%%%%%%%%%
\subsection{Logarithmic operators for general $e+l\in\Z$}
\label{sucsec-logarithmic-op-general}

By using the relations (\ref{vw-recursions}) we may define general special operators. For $e\in{1\over2}\Z$, $\ve\in\Z$ define
\eq$$
J^\ve_e=\Cases{
J^+_{\ve-e}J^+_{\ve+1-e}\cdots J^+_{-1-e},&\text{if $\ve\le0$;}
\\
\ve!^{-1}J^-_{e-\ve}J^-_{e-\ve+1}\cdots J^-_{e-1},&\text{if $\ve>0$.}
}
\label{Jve-e-def}
$$
and $\bar J^\ve_e$ in the same way in terms of $\bar J^\pm_r$. Then define the states
\eq$$
|\Phi^{\circ\ve\bar\ve}_{en}\rangle
=\e^{\i\pi e\bar\ve}J^\ve_e\bar J^{\bar\ve}_e|\Phi^\circ_{e,n+\ve}\rangle,
\qquad
|\Psi^{\ve\bar\ve}_{en}\rangle
=\e^{\i\pi e\bar\ve}J^\ve_e\bar J^{\bar\ve}_e|\Psi_{e,n+\ve}\rangle.
\label{PhiPsi-vebarve-def}
$$
These relations are understood to be valid in the Verma modules.

The corresponding operators $\Phi^{\circ\ve\bar\ve}_{en}(x)$, $\Psi^{\ve\bar\ve}_{en}(x)$ are a primary and a logarithmic operators of conformal dimension
\eq$$
\Delta^{\circ\ve}_{en}=e(e+n)+{|\ve|(|\ve|+1)\over2}.
\label{DeltaO-ve-en-def}
$$

%%%%%%%%%%%%%%%%%%%%%%%%%%%%%%%%%%%%%%%%%%%%%%%%%%%%%%%%%%%%%%%%%%%%%%%%
\section{Conformal blocks and correlation functions at degenerate points}
\label{sec-cb-degen}

In section~\ref{sec-braiding-crossing} we assumed that all values of $e_i$ that enter the conformal blocks, both for operators and for intermediate states, are generic, i.e.\ $e_i+l_i\not\in\Z$. Now we will consider the conformal blocks that contain one degenerate value of $e_i$. First, to complete the identification (\ref{PhiO-def}), we consider special values of $e_i$ on `external' lines of conformal blocks. Then, following the basic idea of Rozansky and Saleur~\cite{Rozansky:1992rx} we turn to the case of specialization in the `internal' lines. We will study the structure of splitting the correlation functions and conformal blocks into degenerate intermediate states.

Everywhere in this section $\cF$ will denote conformal blocks in the Dotsenko\--Fateev normalization. Besides, it will be more convenient to consider the correlation functions in the form $G_{12^*34^*}$ and conformal blocks related to them. Correspondingly, we will assume $L=l_1+l^*_2+l_3+l^*_4$ etc.

First, consider the case, where the state $1$ tends to an atypical one. Our aim is to show that both definitions in (\ref{PhiO-def}) are consistent, i.e.\ the corresponding conformal blocks coincide. Let $e=-l_1$, $e_1-e=\delta\to0$, $n_1=n-{1\over2}$, which corresponds to the first definition. Also let $e'_1=e_1$, $l'_1=l_1+1$, $n'_1=n+{1\over2}$, which corresponds to the second definition. From (\ref{Phi12PhiO3-corr}) we see the channel $s(1)$ does not contribute the correlations functions $G_{12^*34^*}$, since $n+n^*_2+n^*_{s(1)}=1\ne0$. Thus in the case $L=3$ the correlation functions vanish: $G_{12^*34^*}\gamma_{l_2-1}(\delta)\to0$. Analogously, the channel $s(0)$ does not contribute the correlation functions $G_{1'2^*34^*}$, and the correlation functions vanish for $L'=1$. We want to show the identity
\eq$$
\bF_{s(0)}\sbm{2^*&3\\1&4}(z)=\bF_{s(1)}\sbm{2^*&3\\1'&4}(z),\quad\text{if $e_1+l_1=0$,}
\label{bF-e1-degen-id}
$$
which verifies the consistency of the definition (\ref{PhiO-def}) of the operators~$\Phi^\circ_{en}$ and, hence, $\Psi_{en}$ on the level of four\-/point correlation functions.

For $L=2$ in the correlation function $G_{12^*34^*}$ we have the only conformal block
$$
\cF_{s(0)}\sbm{2^*&3\\1&4}(z)\to z^{\delta_{12^*}}(1-z)^{\delta_{2^*3}}I_0(0,1-\ve_2,\ve_3;z)
=\const\cdot z^{\delta_{12^*}}(1-z)^{\delta_{23^*}}\quad\text{as $\delta\to0$.}
$$
With the normalization condition (\ref{bF-norm}), we have in this limit
$$
\bF_{s(0)}\sbm{2^*&3\\1&4}(z)=z^{\delta_{12^*}}(1-z)^{\delta_{23^*}}.
$$
In the correlation function $G_{1'2^*34^*}$ this corresponds to $L'=3$ so that for $\delta=0$ we have
$$
\bF_{s(1)}\sbm{2^*&3\\1'&4}(z)=z^{\delta_{1^{\prime*}2}}(1-z)^{\delta_{23^*}}=z^{\delta_{12^*}}(1-z)^{\delta_{23^*}}=\bF_{s(0)}\sbm{2^*&3\\1&4}(z).
$$
For $L=1$ a less trivial conformal block is for $G_{1'2^*34^*}$:
$$
\cF_{s(1)}\sbm{2^*&3\\1'&4}(z)=z^{\delta_{1'2^*}}(1-z)^{\delta_{2^*3}}I_1(1+\delta,1-\ve_2,\ve_3;z)
=-\delta^{-1}z^{\delta_{1'2^*}+\ve_1-1}(1-z)^{\delta_{2^*3}}+O(\delta^0).
$$
Hence, in the limit $\delta\to0$ we obtain
$$
\bF_{s(1)}\sbm{2^{\prime*}&3\\1&4}(z)=z^{\delta_{1'2^*}+\ve_2-1}(1-z)^{\delta_{2^*3}}=z^{\delta_{12^*}}(1-z)^{\delta_{2^*3}}
=\bF_{s(0)}\sbm{2^*&3\\1&4}(z),
$$
which finishes the proof of~(\ref{bF-e1-degen-id}).

Now turn to the case, where degenerate modules appear in the intermediate states of conformal blocks. The only interesting case here is $L=2$, since the two conformal blocks in the $s$ channel coincide in this case. Let us consider the conformal block $\cF_s\sbm{2^*&3\\1&4}(z)$ with
\eq$$
\ve_{s(\alpha)}\equiv e_1-e_2+l_{s(\alpha)}=k+1-\alpha,
\qquad
k=\ve_1-\ve_2=\ve_4-\ve_3\in\Z.
\label{s-degen-condition}
$$
In this section we associate $s(\alpha)$, $t(\alpha)$, $u(\alpha)$ with the four states $1,2^*,3,4^*$. Consider the $F$ and $B$ matrices in the vicinity of the degenerate point~(\ref{s-degen-condition}). Let
\eq$$
e_{\tilde1}=e_1+\delta,
\qquad
e_{\tilde4}=e_4+\delta,
\label{s-degen-vicinity}
$$
so that $\ve_{\tilde1}=\ve_1+\delta$, $\ve_{\tilde4}=\ve_4+\delta$. Then we have
\Align$$
F\sbm{2^*&3\\\tilde1&\tilde4}
&=F(\ve_1+\delta,1-\ve_2,\ve_3)
={1\over\sin\pi(\ve_2-\ve_3)}\pMatrix{(-1)^k\sin\pi(\ve_2+\delta)&-\sin\pi\ve_2\\(-1)^k\sin\pi(\ve_3+\delta)&-\sin\pi\ve_3},
\label{F-spec}
\\
B^\pm\sbm{2^*&3\\\tilde1&\tilde4}
&=\e^{\pm\i\pi\delta_{2^*3}}B(\ve_1+\delta,1-\ve_2,\ve_3)
\notag
\\*
&={\e^{\pm\i\pi\delta_{2^*3}}\over\sin\pi(\ve_2+\ve_3+\delta)}
\pMatrix{\e^{\mp\i\pi(\ve_3+\delta)}\sin\pi\ve_2&\e^{\pm\i\pi\ve_2}\sin\pi(\ve_2+\delta)\\
  \e^{\mp\i\pi\ve_3}\sin\pi(\ve_3+\delta)&\e^{\pm\i\pi(\ve_2+\delta)}\sin\pi\ve_3}.
\label{B-spec}
$$
At the point $\delta=0$ the matrices are degenerate. Therefore,
\eq$$
\sin^{-1}\pi\ve_2\cdot\cF_{s(0)}\sbm{2^*&3\\1&4}(z)
=\sin^{-1}\pi\ve_3\cdot\cF_{s(1)}\sbm{2^*&3\\1&4}(z)
=\check\cF_s\sbm{2^*&3\\1&4}(z),
\label{cF-s01-identity}
$$
where
\Align$$
\check\cF_s\sbm{2^*&3\\1&4}(z)
&=\sin^{-1}\pi\ve_2\cdot z^{\delta_{22^*}}(1-z)^{\delta_{2^*3}}I_0(\ve_1,1-\ve_2,\ve_3;z)
\notag
\\
&={1\over\sin\pi(\ve_2-\ve_3)}\left((-1)^k\cF_{t(0)}\sbm{2^*&1\\3&4}(1-z)-\cF_{t(1)}\sbm{2^*&1\\3&4}(1-z)\right)
\label{checkcFsO-Ftr}
\\
&={\e^{\mp\i\pi\delta_{2^*3}}\over\sin\pi(\ve_2+\ve_3)}
\left(\e^{\mp\i\pi\ve_3}\cF_{u(0)}\sbm{3&2^*\\1&4}(z^{-1})+\e^{\pm\i\pi\ve_2}\cF_{u(1)}\sbm{3&2^*\\1&4}(z^{-1})\right).
\label{checkcFsO-Btr}
$$
Now consider the transformation inverse to (\ref{F-spec}),~(\ref{B-spec}). We have
\Align$$
F\sbm{2^*&\tilde1\\3&\tilde4}
&=F(\ve_3,1-\ve_2,\ve_1+\delta)
={1\over\sin\pi\delta}
\pMatrix{(-1)^{k+1}\sin\pi\ve_3&(-1)^k\sin\pi\ve_2\\-\sin\pi(\ve_3+\delta)&\sin\pi(\ve_2+\delta)},
\label{Finv-spec}
\\
B^\pm\sbm{3&2^*\\\tilde1&\tilde4}
&=\e^{\pm\i\pi\delta_{2^*3}}B(\ve_1+\delta,\ve_3,1-\ve_2)
={\e^{\pm\i\pi\delta_{2^*3}}\over\sin\pi\delta}
\pMatrix{-\e^{\mp\i\pi(\ve_3+\delta)}\sin\pi\ve_3&\e^{\mp\i\pi\ve_3}\sin\pi(\ve_2+\delta)\\
  \e^{\pm\i\pi\ve_2}\sin\pi(\ve_3+\delta)&-\e^{\pm\i\pi(\ve_2+\delta)}\sin\pi\ve_2}.
\label{Binv-spec}
$$
Though the matrices are divergent, the very transformation has a definite limit in an appropriate basis of conformal blocks. Define
\Align$$
\check\cF_{s'}\sbm{2^*&3\\1&4}(z)
&=\lim_{\delta\to0}{1\over\pi\delta}\left(\sin\pi\ve_2\cdot\cF_{s(1)}\sbm{2^*&3\\\tilde1&\tilde4}(z)
-\sin\pi\ve_3\cdot\cF_{s(0)}\sbm{2^*&3\\\tilde1&\tilde4}(z)\right)
\notag
\\
&={z^{\delta_{12^*}}(1-z)^{\delta_{2^*3}}\over\pi}
\left.{\d\over\d\ve_1}\left(\sin\pi\ve_2\cdot I_1(\ve_1,1-\ve_2,\ve_3;z)
-\sin\pi\ve_3\cdot I_0(\ve_1,1-\ve_2,\ve_3;z)\right)\right|_{\ve_1=\ve_2}.
\label{checcF-s'-def}
$$
Then in the basis $(s,s')$ for $\check\cF$ we have
\Align$$
\cF\sbm{2^*&1\\3&4}(z)
&=\check F\sbm{2^*&1\\3&4}\check\cF\sbm{2^*&3\\1&4}(1-z),
\label{checkcF-Ftr-inv}
\\
\cF\sbm{3&2^*\\1&4}(z)
&=z^{\Delta_4-\sum^3_{i=1}\Delta_i}\check B\sbm{3&2^*\\1&4}^\pm\check\cF\sbm{2^*&3\\1&4}(z^{-1}),
\label{checkcF-Btr-inv}
$$
the matrices $\check F$ and $\check B$ look like:
\Align$$
\check F\sbm{2^*&1\\3&4}
&=\pMatrix{0&(-1)^k\\-\sin\pi(\ve_2-\ve_3)&1},
\label{checkFmat-inv-def}
\\
\check B\sbm{3&2^*\\1&4}^\pm
&=\e^{\pm\i\pi\delta_{2^*3}}
\pMatrix{\e^{\pm\i\pi(\ve_2-\ve_3)}\sin\pi\ve_3&\e^{\mp\i\pi\ve_3}\\
  \e^{\pm\i\pi(\ve_2-\ve_3)}\sin\pi\ve_2&-\e^{\pm\i\pi\ve_2}}.
\label{checkBmat-inv-def}
$$
The inverses to these matrices are
\Align$$
\check F\sbm{2^*&3\\1&4}
&={1\over\sin\pi(\ve_2-\ve_3)}\pMatrix{(-1)^k&-1\\(-1)^k\sin\pi(\ve_2-\ve_3)&0},
\label{checkFmat-def}
\\
\check B\sbm{2^*&3\\1&4}^\pm
&={\e^{\pm\i\pi\delta_{2^*3}}\over\sin\pi(\ve_2+\ve_3)}
\pMatrix{\e^{\mp\i\pi\ve_3}&\e^{\pm\i\pi\ve_2}\\
  \sin\pi\ve_2&-\sin\pi\ve_3}.
  \label{checkBmat-def}
$$
We see that the crossing and braiding transformations in this case are nondegenerate, if we choose an appropriate basis of conformal blocks. Even in the case where there is a degeneration in both $s$ and $t$ channel the transformation is nondegenerate. Indeed, assume
\eq$$
k'=\ve_3-\ve_2=\ve_4-\ve_1\in\Z.
\label{k-tchannel}
$$
Then
\eq$$
\check\cF_s\sbm{2^*&3\\1&4}(z)={1\over\sin\pi\ve_2}\check\cF_{s'}\sbm{2^*&1\\3&4}(1-z).
\label{F11*11-Ftr}
$$
This transformation can be expressed by the matrix
\eq$$
\check F\sbm{2^*&3\\1&4}=\pMatrix{0&\sin^{-1}\pi\ve_2\\\sin\pi\ve_2&0}.
\label{checkFmat-11*11}
$$

Now let us calculate the correlation functions $G_{12^*34^*}$ with
\eq$$
\ve_1-\ve_2=\ve_4-\ve_3=k\in\Z,
\qquad
\bve_1-\bve_2=\bve_4-\bve_3=\bar k\in\Z.
\label{G1234-s-degen}
$$
Hence, we have
\eq$$
e_{s(0)}=e_1-e_2=e_4-e_3=k+l_2-l_1=k+l_3-l_4=\bar k+\bl_2-\bl_1=\bar k+\bl_3-\bl_4.
\label{s(0)-degen}
$$
Not that for $k\ne0$ the contribution (\ref{Ialpha-lim}) in one of the two conformal blocks vanish, so that it is not easy to find the actual leading contribution. It corresponds to the fact that the vector $|w_0^{1+k}\rangle$ for $k<0$ and $|u_0^k\rangle$ for $k>0$ (see Fig.~\ref{fig-felder}) is not at the top of Fock module. Thus the normalization (\ref{bF-norm}) is very tricky in these cases. This is why we will use the Dotsenko\--Fateev normalization, which is continuous at the resonant points.

As for conformal blocks we fist shift $e_1,e_4$ according to~(\ref{s-degen-vicinity}). Then we have
\eq$$
G_{\tilde12^*3\tilde4^*}(z,\bz)=X_{s(0)}\sbm{2^*&3\\\tilde1&\tilde4}
\left(\cF_{s(0)}\sbm{2^*&3\\\tilde1&\tilde4}(z)\bcF_{s(0)}\sbm{2^*&3\\\tilde1&\tilde4}(\bz)
  +{X_{s(1)}\sbm{2^*&3\\\tilde1&\tilde4}\over X_{s(0)}\sbm{2^*&3\\\tilde1&\tilde4}}
  \cF_{s(1)}\sbm{2^*&3\\\tilde1&\tilde4}(z)\bcF_{s(1)}\sbm{2^*&3\\\tilde1&\tilde4}(\bz)\right).
\label{G12*34*-delta}
$$
The prefactor immediately follows from (\ref{X1234-def}):
\eq$$
X_{s(0)}\sbm{2^*&3\\\tilde1&\tilde4}
=(-)^{k-\bar k}\,{\sin\pi\ve_3\sin\pi\bve_3\over\pi^2\delta}X_{12^*34^*}+O(1).
\label{X0-delta}
$$
Note that the square of the factor $X_{12^*34^*}$ is a rational function of $e_2$, $e_3$. The ratio before the second product of conformal blocks is
\eq$$
{X_{s(1)}\sbm{2^*&3\\\tilde1&\tilde4}\over X_{s(0)}\sbm{2^*&3\\\tilde1&\tilde4}}
=-{\sin\pi(\ve_2+\delta)\sin\pi\bve_2\over\sin\pi(\ve_3+\delta)\sin\pi\bve_3}
=-{\sin\pi\ve_2\sin\pi\bve_2\over\sin\pi\ve_3\sin\pi\bve_3}\left(1+\pi\delta\left(\ctg\pi\ve_2-\ctg\pi\ve_3\right)+O(\delta^2)\right).
\label{X1/X0-delta}
$$
In the limit $\delta\to0$ the expression in the parentheses in (\ref{G12*34*-delta}) tends to zero, so that we have to choose the basis $\check\cF_s,\check\cF_{s'}$:
\Multline$$
G_{12^*34^*}(z,\bz)=(-)^{k-\bar k+1}X_{12^*34^*}\Bigl(
  \check\bF_s\sbm{2^*&3\\1&4}(z)\bar{\check\cF}_{s'}\sbm{2^*&3\\1&4}(\bz)+\check\cF_{s'}\sbm{2^*&3\\1&4}(z)\bar{\check\bF}_s\sbm{2^*&3\\1&4}(\bz)
\\*
  +\pi(\ctg\pi\ve_2-\ctg\pi\ve_3)\check\bF_s\sbm{2^*&3\\1&4}(z)\bar{\check\bF}_s\sbm{2^*&3\\1&4}(\bz)
\Bigr),
\label{G12*34*-degenerate}
$$
where
\eq$$
\check\bF_s\sbm{2^*&3\\1&4}(z)=\check\cF_s\sbm{2^*&3\\1&4}(z){\sin\pi\ve_2\sin\pi\ve_3\over\pi}.
\label{checkbF-0-def}
$$

Consider the special case $k=\bar k=0$. In this case we have
\eq$$
\check\bF_s\sbm{2^*&3\\\tilde1&\tilde4}(z)
=\bF_{s(0)}\sbm{2^*&3\\1&4}(z)=\bF_{s(1)}\sbm{2^*&3\\1&4}(z),
\label{checkbF-0}
$$
Besides, define
\Align$$
\check\bF_{s'}\sbm{2^*&3\\\tilde1&\tilde4}(z)
&=\lim_{\delta\to0}\delta^{-1}\left(\bF_{s(1)}\sbm{2^*&3\\\tilde1&\tilde4}(z)-\bF_{s(0)}\sbm{2^*&3\\\tilde1&\tilde4}(z)\right)
\notag
\\
&=\check\cF_{s'}\sbm{2^*&3\\1&4}(z)+(\psi(1-\ve_2)+\psi(\ve_3)-2\psi(1))\check\bF_s\sbm{2^*&3\\1&4}(z).
\label{checkbF-1}
$$
Then the decomposition (\ref{G12*34*-degenerate}) can be rewritten as
\Multline$$
G_{12^*34^*}(z,\bz)=-X_{12^*34^*}\biggl(
  \check\bF_{s'}\sbm{2^*&3\\1&4}(z)\bar{\check\bF}_s\sbm{2^*&3\\1&4}(\bz)
  +\check\bF_s\sbm{2^*&3\\1&4}(z)\bar{\check\bF}_{s'}\sbm{2^*&3\\1&4}(\bz)
\\*
  -(D_2+D_3)\check\bF_s\sbm{2^*&3\\1&4}(z)\bar{\check\bF}_s\sbm{2^*&3\\1&4}(\bz)
\biggr).
\label{G12*34*-k0-degenerate}
$$
This expression generalizes the result of~\cite{Rozansky:1992rx}.

Consider the limit $z\to0$ in this formula. We have
\eq$$
\check\bF_s\sbm{2^*&3\\1&4}(z)=z^{\Delta^\circ_{en}-\Delta_1-\Delta_2}(1+O(z)),
\qquad
\check\bF_{s'}\sbm{2^*&3\\1&4}(z)=-z^{\Delta^\circ_{en}-\Delta_1-\Delta_2}(\log z+O(z)).
\label{checkbF-z->0}
$$
Thus, we have
$$
G_{12^*34^*}(z,\bz)=X_{12^*34^*}z^{\Delta^\circ_{en}-\Delta_1-\Delta_2}\bz^{\Delta^\circ_{en}-\bar\Delta_1-\bar\Delta_2}(\log|z|^2+D_1+D_3+O(|z|)).
$$
It is easy to check that
$$
X_{12^*34^*}=-e^{\i\pi e^2}C^\circ_{s^*12^*}C^\circ_{s34^*}.
$$
Comparing it with (\ref{Psi-Phi-Phi-matel}) and (\ref{Phi12PhiO3-corr}) we obtain
\Multline$$
\Gamma_4^{-1}G_{12^*34^*}(z,\bz)=\langle\Phi_4|\Phi_3(1,1)|\Psi_{en}\rangle\langle\Phi^\circ_{en}|\Phi_{2^*}(z,\bz)|\Phi_1\rangle
+\langle\Phi_4|\Phi_3(1,1)|\Phi^\circ_{en}\rangle\langle\Psi_{en}|\Phi_{2^*}(z,\bz)|\Phi_1\rangle
\\
+O\left(|z|^{2\Delta^\circ_{en}-\Delta_1-\Delta_2+1}\right).
\label{G12*34*-firstterm}
$$

Let us study the decomposition of unity in the $s$ channel of this correlation function from a more general point of view. In the channel we may insert a decompositions of unity. Decompose
\eq$$
\tilde U_N^{-1}=\pMatrix{E^{00}_N&E^{01}_N\\E^{10}_N&E^{11}_N}.
\label{Uinv-decomp}
$$
into the blocks in the same basis $(|\Phi^\circ\rangle,|\Psi\rangle)$. This defines four forms $E^{ij}(e,n)$. Then define
\Align$$
I(e,n)
&=\sum^\sim_{K,\bar K,K',\bar K'}
  \Lambda_{-K}\bar\Lambda_{-\bar K}|\Phi^\circ_{en}\rangle E^{00}_{K\bar K,K'\bar K'}(e,n)\langle\Phi^\circ_{en}|\Lambda_{K'}\bar\Lambda_{\bar K'}
\notag
\\*
&\quad+\sum^\sim_{K,\bar K}\sum_{K',\bar K'}
  \Lambda_{-K}\bar\Lambda_{-\bar K}|\Phi^\circ_{en}\rangle E^{01}_{K\bar K,K'\bar K'}(e,n)\langle\Psi_{en}|\Lambda_{K'}\bar\Lambda_{\bar K'}
\notag
\\*
&\quad+\sum_{K,\bar K}\sum^\sim_{K',\bar K'}
  \Lambda_{-K}\bar\Lambda_{-\bar K}|\Psi_{en}\rangle E^{10}_{K\bar K,K'\bar K'}(e,n)\langle\Phi^\circ_{en}|\Lambda_{K'}\bar\Lambda_{\bar K'}
\notag
\\*
&\quad+\sum_{K,\bar K,K',\bar K'}
  \Lambda_{-K}\bar\Lambda_{-\bar K}|\Psi_{en}\rangle E^{11}_{K\bar K,K'\bar K'}(e,n)\langle\Psi_{en}|\Lambda_{K'}\bar\Lambda_{\bar K'},
\label{Ien-def}
$$
where a tilde over a sum sign means that summation is taken over the reduced basis explained at the end of sec.~\ref{subsec-specmod}. Now the expression
\eq$$
\Gamma_4^{-1}G_{12^*34^*}(z,\bz)
=\langle\Phi_4|\Phi_3(1,1)I(e,n)\Phi_2(z,\bz)|\Phi_1\rangle,
\label{G12*34*-decomp}
$$
provides a decomposition of the four\-/point correlation function. The zero-level contribution with $W^\circ_0=1$, $W'_0=0$ gives~(\ref{G12*34*-firstterm}).

Note that the logarithmic contributions on correlations functions come from the second and the third terms in the decomposition (\ref{Ien-def}). The fourth term only contains the descendants of $|\Psi_{en}\rangle$ and $\langle\Psi_{en}|$ that are not logarithmic. Indeed, by definition (\ref{Uinv-decomp}) we have
$$
(\tilde W^\circ_N\otimes\tilde W^\circ_N)E^{11}_N=0,
\qquad
E^{11}_N(\tilde W^{\circ T}_N\otimes\tilde W^{\circ T}_N)=0.
$$
It means that the vectors of the form $\Lambda_{-K}\bar\Lambda_{-K'}|\Phi^\circ_{en}\rangle$, which are the only terms in $\Phi_2(z,\bz)|\Phi_1\rangle$ that have logarithmic coefficients do not contribute the fourth term. The vectors of the form $\Lambda_{-K}\bar\Lambda_{-K'}|\Psi_{en}\rangle$ always enter with the coefficients proportional to powers of $z,\bz$. Consistently, the vectors $\langle\Phi^\circ_{en}|\Lambda_K\bar\Lambda_{K'}$ from the decomposition of $\langle\Phi_4|\Phi_3(1,1)$ do not enter the fourth term either.

%%%%%%%%%%%%%%%%%%%%%%%%%%%%%%%%%%%%%%%%%%%%%%%%%%%%%%%%%%%%%%%%%%%%%%%%
\section{Discussion}
\label{sec-discussion}

In the revision of free field realization of $GL(1|1)$ WZW model undertaken in this paper we succeeded to fix (up to a sign) all the structure constants needed for crossing symmetry of right\-/left symmetric four\-/point functions with only typical modules representatives involved. An important tool used in this paper was splitting of the representations of the $\widehat{gl}(1|1)$ algebra into representations of the algebra of neutral currents ($T,J^N,J^E$) according to the fermion number. In particular, this splitting shows that the conformal theory consists of the Neveu\--Schwartz and Ramond sectors.

The Moore\--Seiberg identities for the braiding and crossing matrices of the conformal blocks are satisfied when the intermediate states also correspond to typical representations. The atypical representation in the intermediate states correspond to degeneration of the braiding and crossing matrices due to the degeneration of the standard basis of conformal blocks. This degeneration can be resolved following the ideas of Rozansky and Saleur, which leads to well defined three\-/{} and four\-/point correlation functions. We see that the normalization of conformal blocks related to the free field representation is more relevant here than the canonical normalization. In this normalization the conformal blocks continuously cross the special points where the intermediate states degenerate, while the canonically normalized conformal blocks are generally discontinuous in the parameters of the operators.

\section{Acknowledgments}
\label{sec-acknowledgments}

I am grateful to A.~Babichenko, who introduced me into the problems of the $gl(1|1)$ model, M.~Bershtein and A.~Litvinov for discussions. The work was supported the Russian Science Foundation under the grant 18--12--00439.

%\raggedright
%\bibliographystyle{utphys}
%\bibliography{main}

\providecommand{\href}[2]{#2}\begingroup\raggedright\endgroup

\end{document}